\documentclass[aps,prb,twocolumn]{revtex4}  

\usepackage{graphicx}  
\usepackage{epstopdf}
\usepackage{dcolumn}   
\usepackage{bm}        
\usepackage{amsmath}
\usepackage{amssymb}   
\usepackage{array}
\usepackage[caption=false]{subfig}
\usepackage[bookmarks=false,linkcolor=blue,urlcolor=blue,colorlinks,citecolor=blue]{hyperref}
\usepackage{exscale,relsize}
\usepackage{xfrac}
\usepackage{cleveref}
\usepackage[bookmarks=false,linkcolor=blue,urlcolor=blue,colorlinks,citecolor=blue]{hyperref}

\global\long\def\bra#1{\left\langle #1\right|}
\global\long\def\ket#1{\left|#1\right\rangle }
\global\long\def\braket#1#2{\left\langle #1|#2\right\rangle }

\global\long\def\dag{^{\dagger}}

\global\long\def\tens{\otimes}

\global\long\def\eps{\epsilon}

\global\long\def\inf{\infty}

\global\long\def\s{\sigma}

\global\long\def\a{\alpha}

\global\long\def\P{\mathcal{P}}

\def \tx#1{{\rm#1}}

\makeatletter

\begin{document}

\title{Realizing Topological Superconductivity with Superlattices}
\author{Yoav Levine, Arbel Haim, and Yuval Oreg}
\affiliation{Department of Condensed Matter Physics, Weizmann Institute of Science,
Rehovot 7610001, Israel}

\date{\today}

\begin{abstract}
The realization of topological superconductors (SCs) in one or two dimensions is a highly pursued goal. Prominent proposed realization schemes include semiconductor/superconductor heterostructures and set stringent constraints on the chemical potential of the system. However, the ability to keep the chemical potential in the required range while in the presence of an adjacent SC and its accompanied screening effects, is a great experimental challenge.
In this work, we study a SC lattice structure in which the SC is deposited periodically on a one- or two-dimensional sample. We demonstrate that 
this realization platform overcomes the challenge of controlling the chemical potential in the presence of the superconductor's electrostatic screening. We show how Majorana bound states emerge at the ends of a one-dimensional system proximity coupled to a one-dimensional SC lattice, and move on to present a SC-lattice-based realization of the two-dimensional $p_{x}+ip_{y}$ SC, hosting chiral Majorana modes at its edges. In particular, we establish that even when assuming the worst case of absolute screening, in which the chemical potential under the SC is completely unaffected by the external gate potential, the topological phase can be reached by tuning the chemical potential in the area not covered by the SC. Finally, we briefly discuss possible effects of Coulomb blockade on the properties of the system.
\end{abstract}

\maketitle

\section{Introduction\label{sec:Intro}}

Topological superconductors (SCs) have been a topic of great interest
in recent years \cite{0034-4885-75-7-076501,Beenakker2013search}.
Promising proposals for realizing topological SCs often involve proximity
coupling a SC to a metallic or semiconducting system \cite{PhysRevLett.100.096407,PhysRevB.79.161408,Sau2010,PhysRevB.81.125318,Duckheim2011andreev,PhysRevLett.105.077001,PhysRevLett.105.177002}.
In most proposals, the chemical potential of the system is required
to reside in a specific and narrow range in order to facilitate the
topological phase. This requirement poses a great experimental challenge. In particular, the SC screens electric fields in the normal system, making it difficult to tune the chemical potential by using a gate voltage.

A prominent proposal for realizing a topological SC, in both one~\cite{PhysRevLett.105.077001,PhysRevLett.105.177002} and two~\cite{Sau2010,PhysRevB.81.125318} dimensions, involves a semiconductor with strong spin-orbit coupling (SOC), subjected to a magnetic field and in proximity to a conventional $s$-wave SC. In one dimension (1d), several groups have observed signatures of topological superconductivity in such systems~\cite{Mourik1003,Das2012,Deng2012Anomalous,Rokhinson2012,PhysRevLett.110.126406,PhysRevB.87.241401}.
The challenge of tuning the chemical potential was initially addressed by covering one side of
the semiconducting wire with the SC, while placing a gate across the other side. This has enabled some control over the electron density under the SC contact. More recently, another
approach was employed, in which the SC was grown as an island on top of the semiconductor
\cite{albrecht2016exponential}. This reduces the electric screening by the SC, due to the fact that the SC is floating rather than grounded~\cite{FloatingSuperconductorComment}, and introduces Coulomb blockade effects. Signatures of topological superconductivity have also been observed~\cite{Nadj-Perge602,Pawlak2016probing,Ruby2015end} in ferromagnetic atomic chains on a superconducting substrate~\cite{Nadj-Perge2013proposal,Braunecker2013interplay,Vazifeh2013self,Klinovaja2013topological,Pientka2013Topological,Brydon2015topological,Peng2015strong,Dumitrescu2015majorana}.

To the best of our knowledge, the two-dimensional (2d) version of the above proposal~\cite{Sau2010,PhysRevB.81.125318} has not been experimentally realized to date. The ability to perform successful gating and to tune the chemical potential to the range required for the topological phase, is yet to be demonstrated in 2d systems.

In this paper, we study the possibility of overcoming the challenge of gating by creating a structure where the SC is deposited periodically on the sample, forming a superconducting lattice, as is depicted in Fig. \ref{fig:Setup}. In such a setup, either in 1d or in 2d, the electrostatic screening caused by the SC occurs only in certain spatial regions, where in the other regions (those not covered by a SC) the chemical potential can be more easily tuned, by using a gate voltage. In 1d, previous works have studied related periodic structures~\cite{sau2012realizing,sau2012avoidance,fulga2013adaptive,malard2016synthesizing,hoffman2016topological,zhang2016majorana,PhysRevB.94.024507}. In 2d, a periodic structure of alternating SC/Ferromagnetic segments coupled to a 2DEG was also suggested~\cite{Lee2012electrical} as a possible realization of a topological SC.

In 1d, we show how Majorana Bound States (MBSs) appear at the ends of a system coupled to a 1d SC lattice [cf. Fig.~\hyperref[fig:Setup]{\ref{fig:Setup}(a)}]. We analyze different regimes of the ratio between the SC lattice unit cell and other relevant length scales in the problem, such as Fermi wavelength, superconducting coherence length, and SOC length. As these length scales are typically of the same order, they provide a convenient reference. We demonstrate that choosing the length of the lattice unit cell comparable to these length scales provides enhanced accessibility to the topological regions in parameter space.

In the 2d case, we show that chiral Majorana modes appear at the edges of a system coupled to a 2d SC lattice [cf. Fig.~\hyperref[fig:Setup]{\ref{fig:Setup}(b)}]. Concentrating on the experimentally-relevant case, in which the SC lattice constant is of the order of the other length scales in the system, we show that the SC lattice platform allows to tune into the topological phase by controlling the chemical potential in the normal non-SC regions. Importantly, for both the 1d and 2d cases, we find that even when assuming the extreme case of absolute screening, in which the chemical potential under the SC is completely unaffected by the external gate potential, the topological phase can be reached by controlling the chemical potential in the regions lacking proximity to the SC.

Our results are relevant to strong-spin-orbit-coupled semiconductors, such as InAs or InSb, in proximity to a superconductor (for example Al). However, we further demonstrate that our conclusions hold also for a scenario in which the SC, but not the semiconductor, has strong SOC. This has relevance to systems composed of light chemical elements, such as Carbon nanotubes (in 1d) or Graphene (in 2d), covered by a heavy-element SC such as NbN or Pb.

In the present work, we focus on systems which lack disorder. While strong disorder can certainly change some of our conclusions, the topological properties of the system should hold in the presence of weak disorder~\cite{Motrunich2001Griffiths,Brouwer2011Probability}. In particular, one expects that as long as the energy gap is large in comparison to $1/\tau$, where $\tau$ is a time-scale associated with disorder, the zero-energy MBSs at the system's ends would remain protected.

The rest of this paper is organized as follows. In Sec.~\ref{sec:1D-superconducting-lattice} we discuss the realization of the 1d topological phase using a SC lattice structure, analyzing different parameter regimes. In Sec.~\ref{sec:2D_phase} we extend our results to two dimensions. We conclude in Sec.~\ref{sec:discussion} and discuss methods of experimentally realizing the suggested SC lattice setup and possible effects of charging energy on our system, including the stabilization of topologically-ordered phases by Coulomb-blockade effects. An appendix with a few details of the calculations is included.

\section{\label{sec:1D-superconducting-lattice}a one-dimensional superconducting lattice}

\subsection{The Model\label{sec:1D_model}}

\begin{figure}
\begin{tabular}{cc}
\includegraphics[scale=0.185]{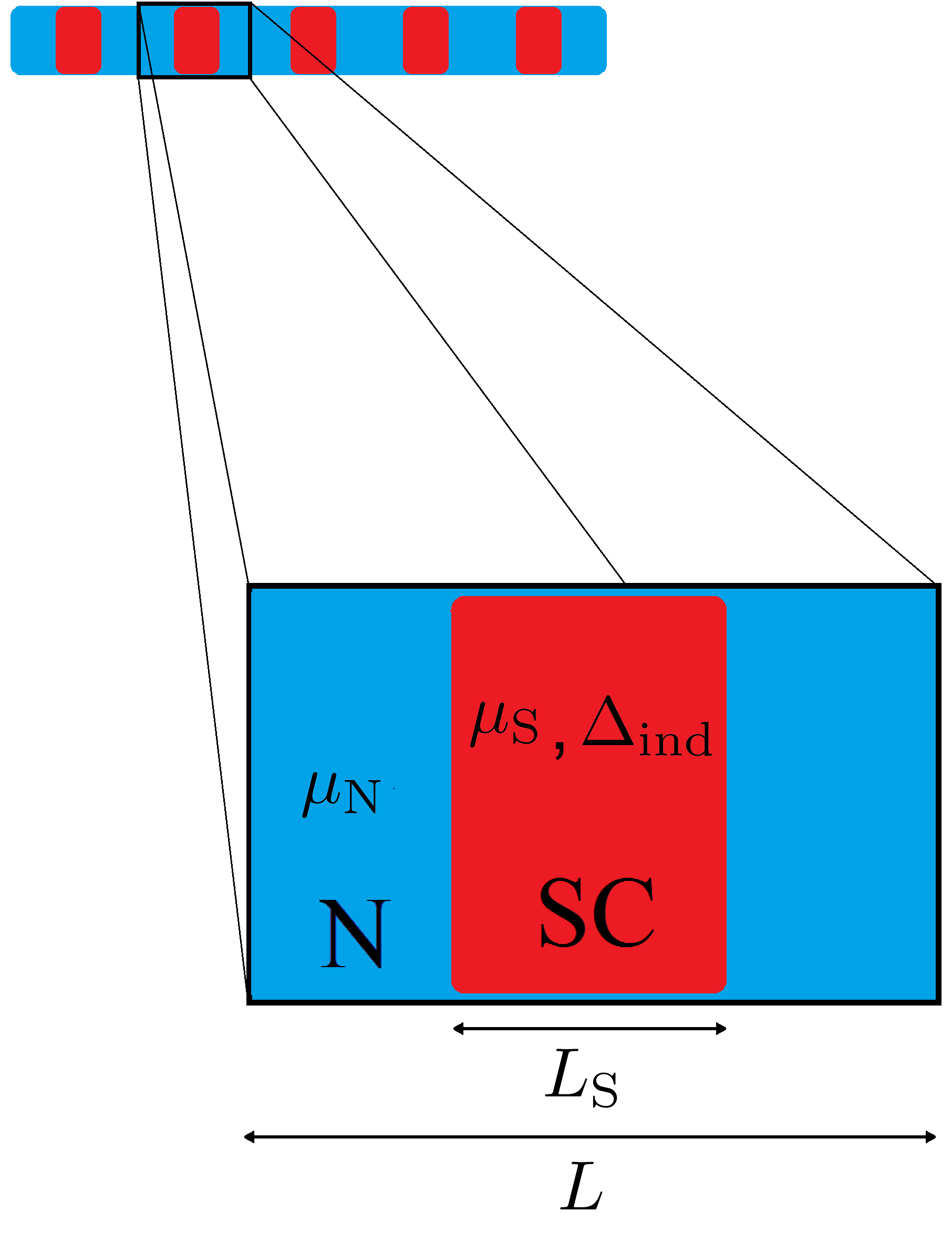}
\llap{\parbox[c]{7cm}{\vspace{-6cm}\footnotesize{(a)}}}

& \hskip 1mm

\includegraphics[scale=0.185]{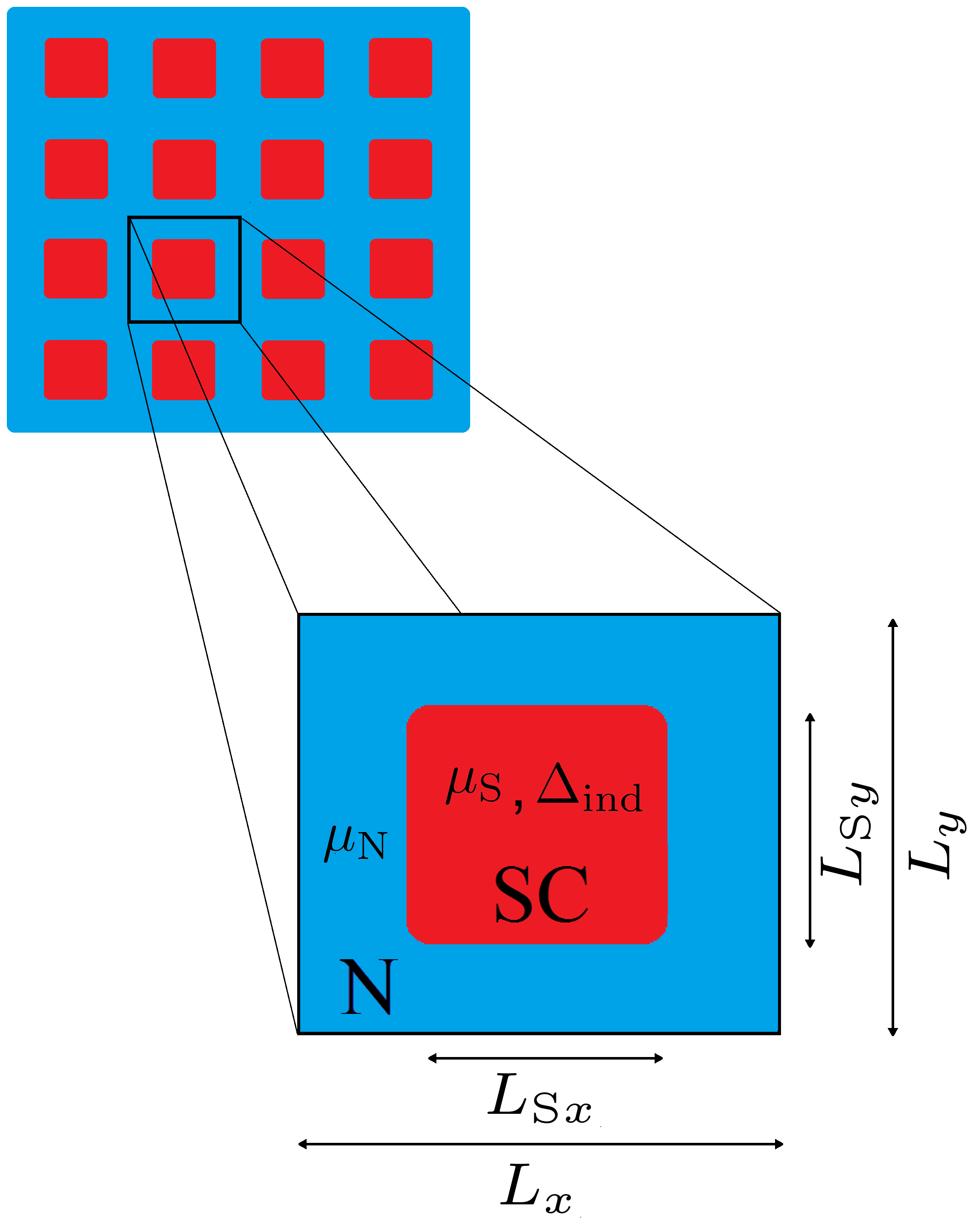}
\llap{\parbox[c]{9.5cm}{\vspace{-6cm}\footnotesize{(b)}}}

\end{tabular}
\caption{A schematic illustration of a system covered by a SC lattice in (a) 1d and (b) 2d. Under each system is an enlarged picture of the unit cell composing the super-lattice, where lengths and quantities are marked. Red color marks regions with induced SC due to contact with a SC, while blue marks normal regions not in proximity to a SC. The red regions have an induced pairing of $\Delta_{\rm ind}$ and a chemical potential $\mu_{\rm S}$. The blue regions have no induced pairing term, and a chemical potential $\mu_{\rm N}$. The chemical potential values to be used for $\mu_{\rm S}$ and $\mu_{\rm N}$ are given by the distance between the bottom of the conduction band in the respective regions to the general electro-chemical potential, which is uniform in equilibrium.\label{fig:Setup}}
\end{figure}

As noted above, we address the challenge of controlling the chemical potential by considering a system composed of a SC which is periodically deposited on a semiconductor nanowire in a super-lattice geometry, as depicted in Fig.~\hyperref[fig:Setup]{\ref{fig:Setup}(a)}. In the presence of an external magnetic field, the Hamiltonian of the system reads
\begin{equation}
\begin{split}
&H=\int \tx{d}x \Psi^\dagger(x) \mathcal{H}(x) \Psi(x)~;\\
&\mathcal{H}(x) = \left[-\partial^2_x/2m^\ast-\mu(x)\right]\tau_z+i\alpha\tau_z\sigma_y\partial_{x} + V_{\rm Z}\sigma_x \\
& \hskip 8mm + \Delta(x)\tau_x,
\end{split}\label{eq:1D real space hamiltonian}
\end{equation}
with $\Psi^\dagger(x) = [\psi^\dagger_\uparrow(x),\psi^\dagger_\downarrow(x),\psi_\downarrow(x),-\psi_\uparrow(x)]$, and where $m^\ast$ is the effective electron mass, $\a$ is the Rashba SOC
constant, $V_{\rm Z}$ is the Zeeman coupling due to the applied magnetic
field, $\mu\left(x\right)$ is the chemical potential in the wire,
and $\Delta\left(x\right)$ is the proximity-induced pairing
potential. Here, $\s_j$ and $\tau_j$ with $j\in\{x,y,z\}$ are sets of Pauli matrices acting on the spin and particle-hole degrees of freedom, respectively. We set $\hbar$ =1 throughout the paper.

Before analyzing the spatially inhomogeneous Hamiltonian of Eq.~\eqref{eq:1D real space hamiltonian}, we note that for the uniform case, $\mu\left(x\right)=\mu_{\rm 0}$,
$\Delta\left(x\right)=\Delta_{\rm ind}$, this system has been shown to be in a topological superconducting phase whenever~\cite{PhysRevLett.105.077001,PhysRevLett.105.177002}
\begin{equation}
|\mu_{\rm 0}|<\sqrt{V_{\rm Z}^{2}-\Delta_{\rm ind}^{2}}\equiv\mu_{\rm c}.\label{eq:uniform condition 1D}
\end{equation}
In typical experimental setups, the chemical potential of the semiconducting
wire will be much higher than $\mu_{\rm c}$, due to the work-function difference between the SC and the semiconducting wire. In the absence of control over the chemical potential (which is compromised by the electrostatic screening of the SC) the system will therefore remain topologically trivial.

The situation changes when one considers a spatially-modulated superconducting proximity coupling, as depicted in Fig.~\hyperref[fig:Setup]{\ref{fig:Setup}(a)}. The induced pair potential is given in this case by
\begin{equation}\label{eq:Delta_modulation}
\begin{split}
&\Delta(x) = \Delta_{\rm ind} \sum_{n\in\mathbb{Z}} {\rm rect}\left(\frac{x-nL}{L_{\rm S}}\right);\\
&{\rm rect}(s) = \left\{
\begin{array}{lr}
1,&|s|<0.5\\
0,&|s|\ge 0.5
\end{array}
\right.,
\end{split}
\end{equation}
where $L_{\rm S}$ is the length of each superconducting segment and $L$ is the period of the modulation. As a result of the screening by the deposited SC, the chemical potential also becomes spatially modulated, and can be described by
\begin{equation}\label{eq:mu_modulation}
\mu(x) = \mu_{\rm N} + (\mu_{\rm S}-\mu_{\rm N}) \sum_{n\in\mathbb{Z}} {\rm rect}\left(\frac{x-nL}{L_{\rm S}}\right),
\end{equation}
where $\mu_{\rm S}$ ($\mu_{\rm N}$) is the chemical potential in the parts of the wire (not) covered by the SC. We note that due to the spread of the wave function across the wire, the transition between $\mu_N$ and $\mu_S$
may be smoother, an effect which we have not included in the present calculation. Our conclusions, however, do not rely on the exact form of the chemical potential's modulation. In what follows, we assume that while $\mu_{\rm S}$ is fixed for a given experimental setup, $\mu_{\rm N}$ can be controlled using a bottom gate voltage.

We are interested in the conditions under which the system of Eq.~\eqref{eq:1D real space hamiltonian} is in the topological superconducting phase, hosting a pair of zero-energy MBSs at its ends. Also of interest is the behavior of the excitation gap, which in turn protects these MBSs, as a function of the system's parameters.

To obtain a criterion for when the system is in the topological phase, it is convenient to impose periodic boundary conditions and work in momentum space. By inserting the Fourier expansions $\mu(x) = \sum_n \mu_n\exp(ixG_n)$ and $\Delta(x) = \sum_n \Delta_n\exp(ixG_n)$ in Eq.~\eqref{eq:1D real space hamiltonian}, one obtains
\begin{equation}
\begin{split}
&H = \frac{1}{2}\int_{-\pi/L}^{\pi/L} {\rm d}k \sum_{m,n} \Psi^\dagger(k+G_m)\mathcal{H}^{\rm B}_{mn}(k) \Psi(k+G_{n}),
\end{split}
\end{equation}
where $G_n = 2\pi n/L$, and with the Bloch Hamiltonian given by
\begin{equation}\label{eq:Bloch_H}
\begin{split}
\mathcal{H}^{\rm B}_{mn} &=
\left\{ \left[\frac{(k+G_m)^2}{2m^\ast}  + \alpha(k+G_m)\sigma_y\right]\tau_z + B\sigma_x \right\}\delta_{mn}\\
&+ \sum_{n'} \left(-\mu_{n'} \tau_z + \Delta_{n'} \tau_x\right)\delta_{m+n',n}.
\end{split}
\end{equation}
For the modulation considered in this work, specified by Eqs.~\eqref{eq:Delta_modulation} and \eqref{eq:mu_modulation}, one has
\begin{equation}\label{eq:Fourier_expan}
\Delta_n = \Delta_{\rm ind}\frac{\sin(\pi nr)}{\pi n} \hskip 2mm ; \hskip 2mm
\mu_n = \mu_{\rm N}\delta_{n,0} + (\mu_{\rm S} - \mu_{\rm N})\frac{\sin(\pi nr)}{\pi n},
\end{equation}
where $r\equiv L_{\rm s}/L$ is the relative part of the wire which is covered by a superconductor.

The topological classification of a Hamiltonian depends on the anti-unitary symmetries that it obeys. In our case, one can check that the Hamiltonian of Eq.~\eqref{eq:Bloch_H} obeys a particle-hole symmetry
\begin{equation}
\Lambda{\mathcal{H}^{\rm B}}^\ast(-k)\Lambda^{-1} = -\mathcal{H}^{\rm B}(k) \hskip 2mm ; \hskip 2mm
\Lambda_{mn} = \tau_y\sigma_y\delta_{m,-n},
\end{equation}
and a time-reversal symmetry
\begin{equation}\label{eq:TRS}
T{\mathcal{H}^{\rm B}}^\ast(-k)T^{-1} = \mathcal{H}^{\rm B}(k) \hskip 2mm ; \hskip 2mm
T_{mn} = \delta_{m,-n},
\end{equation}
which both squares to $+1$, namely $TT^\ast=1$ and $\Lambda\Lambda^\ast=1$. This places the Hamiltonian $\mathcal{H}^{\rm B}$ in symmetry class BDI~\cite{PhysRevB.55.1142}, with a $\mathbb{Z}$ topological invariant~\cite{PhysRevB.78.195125,kitaev2009periodic}, which can support an integer number of zero-energy MBSs at each end. Upon breaking the time-reversal symmetry of Eq.~\eqref{eq:TRS}, the system is in symmetry class D with a $\mathbb{Z}_2$ invariant, which can support a \emph{single} MBS at each end. Since this time-reversal symmetry is in practice rather fragile~\cite{FragileSymmetry}, we are primarily interested in the class D topological invariant~\cite{1063-7869-44-10S-S29,ghosh2010non},
\begin{equation}
\mathcal{Q}={\rm sgn}\left\{ {\rm Pf}\left[\Lambda \mathcal{H}^{\rm B}(k=0)\right]\right\} {\rm sgn}\left\{ {\rm Pf}\left[\Lambda \mathcal{H}^{\rm B}(k=\pi/L)\right]\right\} .\label{eq:Kiteav invariant}
\end{equation}
This invariant was shown~\cite{tewari2012topological} to be equivalent to the parity of the class BDI $\mathbb{Z}$ topological invariant. Indeed, upon coupling an odd number of MBSs [by breaking the time-reversal symmetry of Eq.~\eqref{eq:TRS}], a single protected MBS remains.

\subsection{The Short-Lattice-Constant Regime}\label{sec:1dresults}

\begin{figure}
\begin{tabular}{cc}
\includegraphics[clip=true,trim =0mm 0mm 0mm 0mm,width=0.23\textwidth]{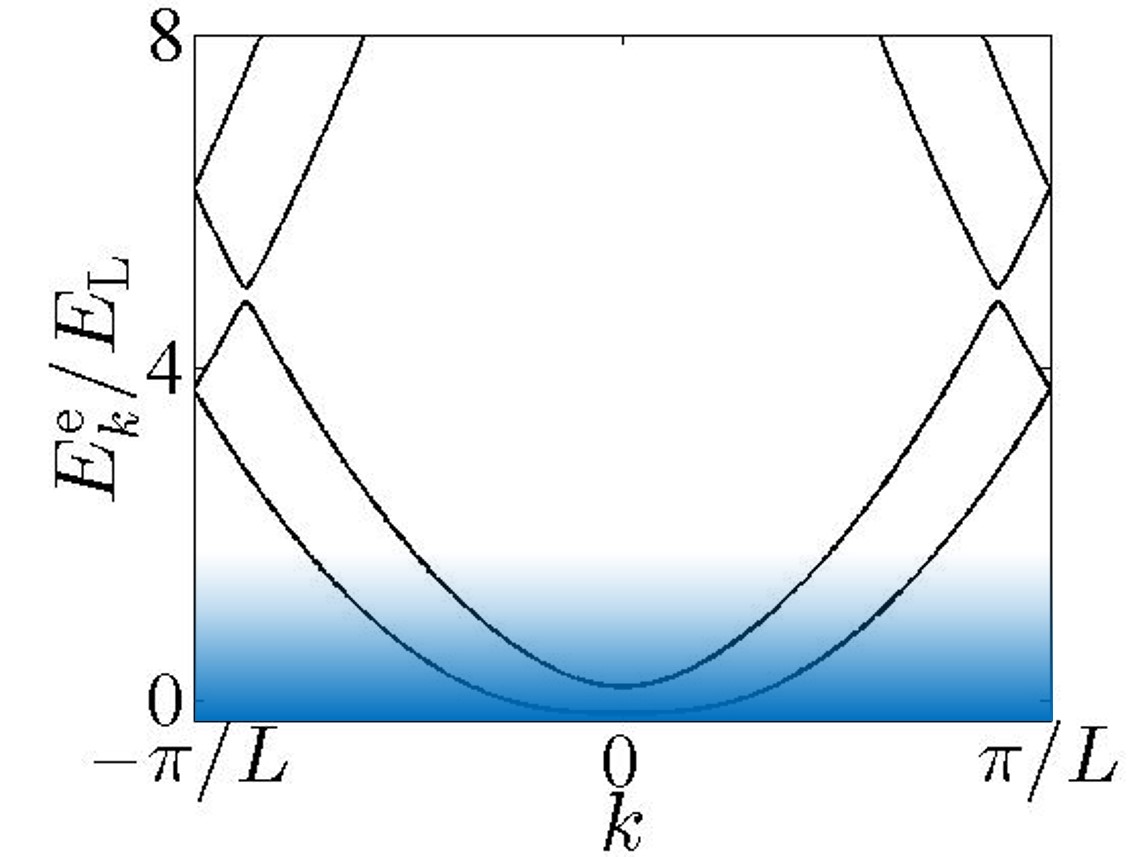}
\llap{\parbox[c]{0.45\textwidth}{\vspace{-0.35\textwidth}{(a)}}}
&
\includegraphics[clip=true,trim =0mm 0mm 0mm 0mm,width=0.23\textwidth]{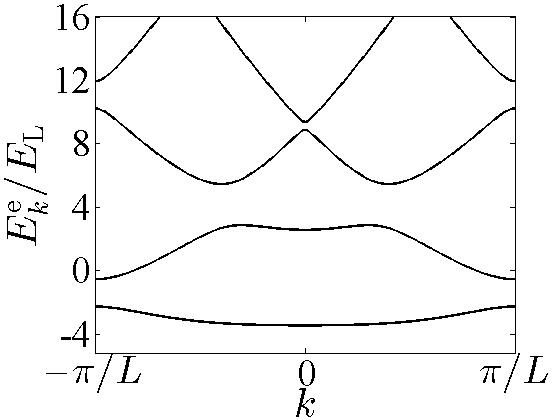}
\llap{\parbox[c]{0.45\textwidth}{\vspace{-0.35\textwidth}{(b)}}}
\end{tabular}
\caption{Electron dispersion of the Bloch Hamiltonian, Eq.~\eqref{eq:Bloch_H} (namely excluding the holes part of the spectrum and setting $\Delta_n=0$). (a) An example of the short-lattice-constant regime where the length of the unit cell, $L$, is the shortest length scale in the problem. Alternatively stated, $E_{\rm L}\equiv 1/(m^\ast L^2)$ is larger than all other energy scales in the problem. In this limit, the system is approximately described by the Hamiltonian of a uniform system with the average chemical potential and average induced pairing potential. The blue shaded area represents the range of the average chemical potential, $\mu_0$, where it is sufficiently small compared to $E_{\rm L}$, such that the above approximation holds. Here, we have used $V_{\rm Z}=0.16E_{\rm L}$, $E_{\rm so}=0.07E_{\rm L}$, $\mu_0=0$, and $\mu_1=0.09E_{\rm L}$ [see Eq.~\eqref{eq:Fourier_expan}]. (b) An example of the electron part of the spectrum for a larger lattice constant, $L$, giving rise to $V_{\rm Z}=3E_{\rm L}$, $E_{\rm so}=1.4E_{\rm L}$, $\mu_0=0$, and $\mu_1=1.7E_{\rm L}$. Here, the band structure and the exact location of the chemical potential may play an important role.\label{fig:minibands}}
\end{figure}

We begin by examining the limit where the length of the unit cell, $L$, is short enough such that $E_{\rm L}\equiv1/(m^\ast L^2)$ is much larger than all other relevant energy scales, namely $\mu_{\rm N}$, $\mu_{\rm S}$, $V_{\rm Z}$, $\Delta_0$, and $E_{\rm so}=m^\ast\alpha^2/2$. We refer to this limit as the short-lattice-constant regime. Fig.~\hyperref[fig:minibands]{\ref{fig:minibands}(a)} depicts an example of the electronic dispersion in this regime.

\begin{figure*}
\begin{tabular}{cc}

\includegraphics[scale=0.5]{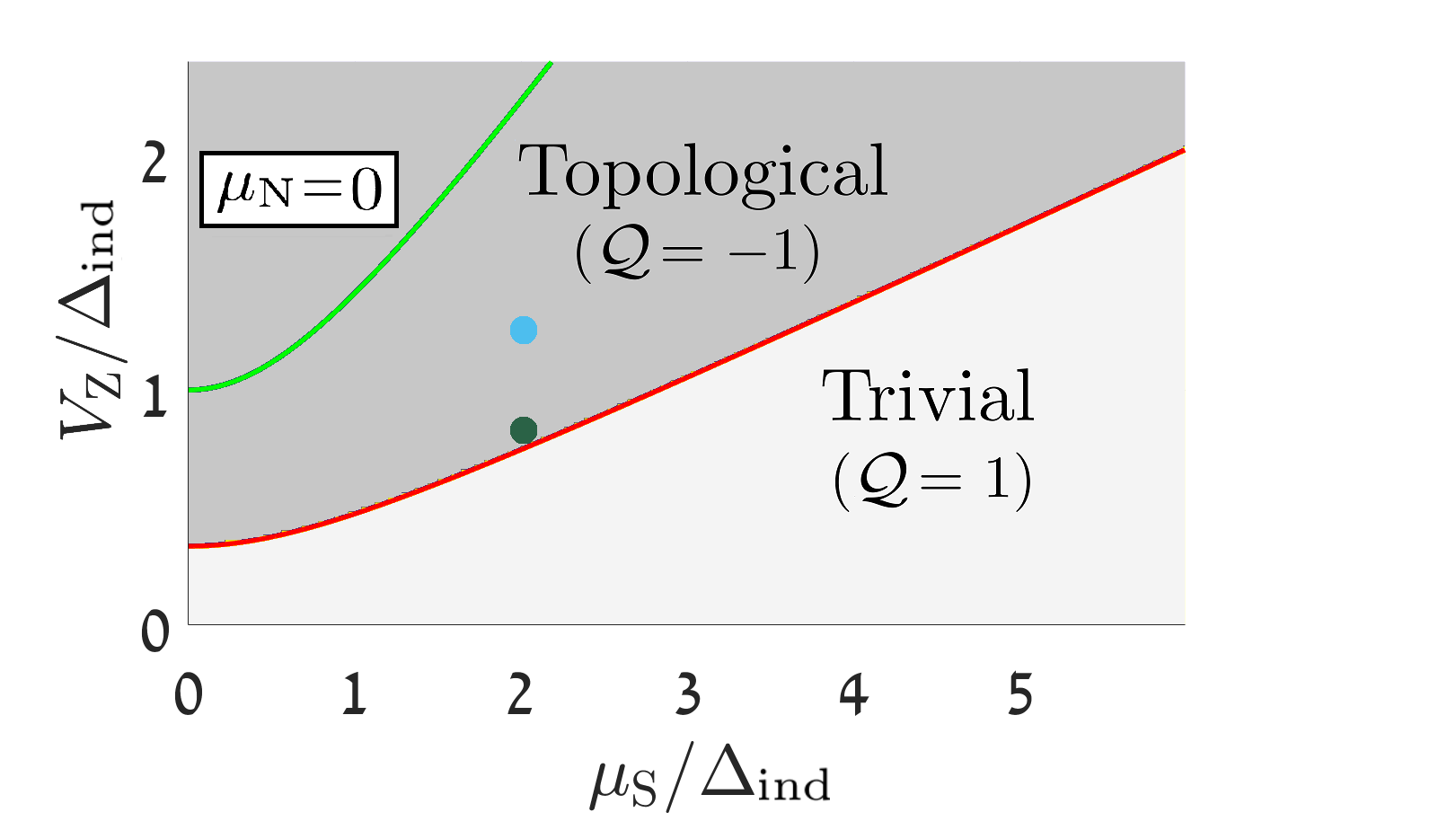}
\llap{\parbox[c]{14.38cm}{\vspace{-8.3cm}{(a)}}}
\llap{\parbox[c]{10.7cm}{\vspace{-5.43cm}\footnotesize{A}}}
\llap{\parbox[c]{10.9cm}{\vspace{-4.2cm}\footnotesize{B}}}
& \hskip 1mm

\includegraphics[scale=0.47]{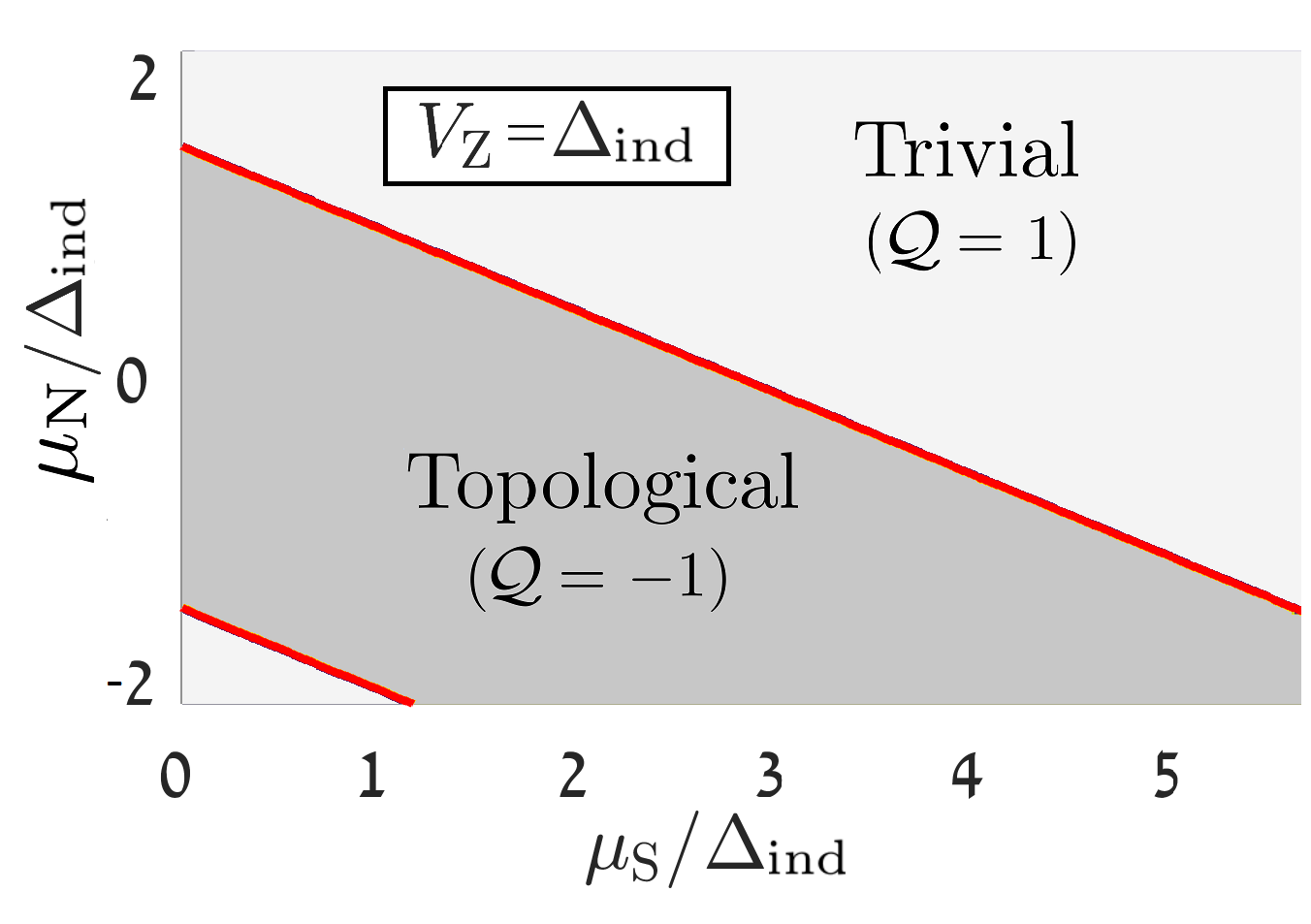}
\llap{\parbox[c]{11cm}{\vspace{-8.33cm}{(b)}}}
 \\ \hskip -4.2mm

\includegraphics[scale=0.5]{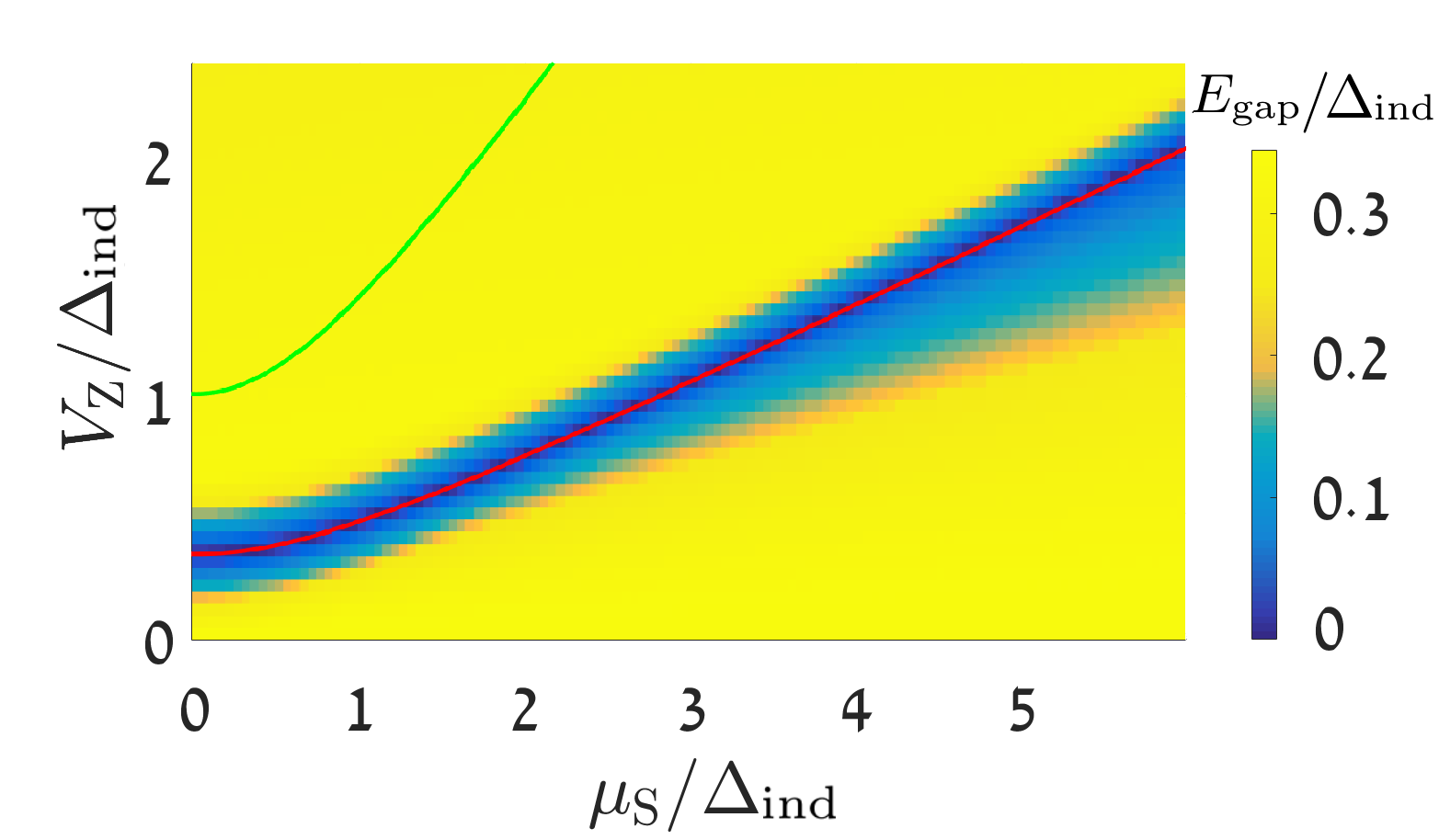}
\llap{\parbox[c]{13.94cm}{\vspace{-8.27cm}{(c)}}}

 &
\includegraphics[scale=0.53]{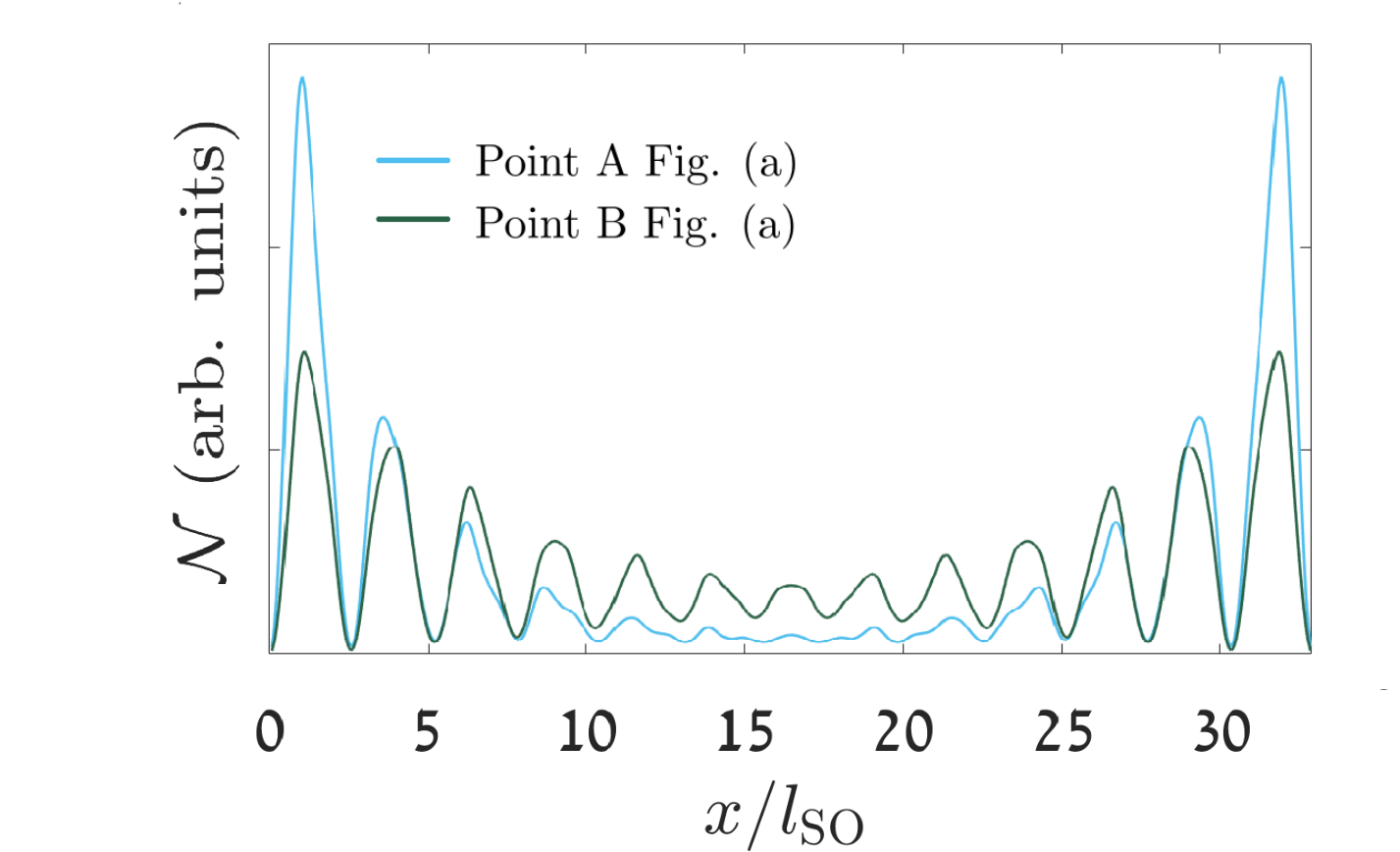}
\llap{\parbox[c]{12.1cm}{\vspace{-8.45cm}{(d)}}}
\end{tabular}
\caption{Numerical simulations of the system described in Eqs.~\eqref{eq:1D real space hamiltonian},~\eqref{eq:Delta_modulation} and~\eqref{eq:mu_modulation} and depicted in Fig.~\hyperref[fig:Setup]{\ref{fig:Setup}(a)}, in the short-lattice-constant regime, where $E_\tx{L}=1/(m^\ast L)$ is the leading energy scale. The parameters used in the simulation are compatible with an InAs nanowire in proximity to superconducting Al~\cite{Das2012}, namely $E_{\rm so}=75\mu{\rm eV}$, $l_{\rm so}=130{\rm nm}$, and $\Delta_{\rm ind}=50\mu{\rm eV}$. The length of the superlattice unit cell is $L=30\tx{nm}$, corresponding to $E_\tx{L}=2.8\tx{meV}$, and the part of the unit cell covered with a SC is $r=L_\tx{S}/L=1/3$.
(a) The topological invariant, $\mathcal{Q}$, as a function of the Zeeman field,
$V_{\rm Z}$, and the chemical potential in the SC regions, $\mu_{\rm S}$,
calculated according to Eq. \eqref{eq:Kiteav invariant}. The chemical potential in the normal region is fixed at $\mu_\tx{N}=0$.
The topological ($\mathcal{Q}=-1$) region in parameter space is marked in dark grey and the trivial ($\mathcal{Q}=1$) region in parameter space is marked in light grey. As expected in the short-lattice-constant regime, the phase transition follows that of a uniform system with modified effective parameters, $\mu_0=r\mu_{\rm S}$ and $\Delta_0=r\Delta_\tx{ind}$, as described by Eq. \eqref{eq:fast_mod_phase_bound_muN0} which is marked by the red line. The green line follows Eq. \eqref{eq:uniform condition 1D} and represents the phase transition line for a case where the system is uniformly covered by the SC. An extension of the topological region into lower zeeman fields and higher values of the chemical potential due to the SC lattice is observed. (b) The topological invariant, $\mathcal{Q}$, as a function of the chemical potential in the normal region, $\mu_{\rm N}$, and the chemical potential in the SC regions, $\mu_{\rm S}$, for a fixed Zeeman field, $V_\tx{Z} = \Delta_\tx{ind}=50\mu\tx{eV}$.
The phase transition line clearly follows Eq. \eqref{eq:muN muS L_<<_Lambda condiiton}, which is marked by the red line. Notice that in this limit $\mu_\tx{N}$, which is the experimentally controlled parameter, can be used to tune the system into the topological phase for any value of $\mu_\tx{S}$.
(c) The excitation energy gap normalized by $\Delta_\tx{ind}$. As expected, the gap closes at the phase transition between the trivial and topological phases. In a uniformly-covered system, the topological gap for Zeeman fields which are far way from the critical line but still small enough compared with $2E_{\rm so}$, is approximately given by $E_{\rm gap}\simeq \Delta_\tx{ind}$~\cite{GapRemark}. Here, a reduction of the gap is observed, $E_\tx{gap}\simeq r\Delta_\tx{ind}$, due to the SC only partially covering the nanowire. (d) The local density of states (LDOS) in arbitrary units for the lowest-energy excitation, calculated according to Eq. \eqref{eq:LDOS} in Appendix~\ref{app:ldos}. A tight-binding simulation was used, of a finite wire of length $30 l_{\rm so} $.
The LDOS is shown for two points marked in (a). Light blue represents point A: $(\mu_{\rm S} = 2\Delta_\tx{ind},V_{\rm Z} = 1.2\Delta_\tx{ind})$, located well above the phase transition line where the gap is approximately $\Delta_0=r\Delta_\tx{ind}$, while dark green is for point B: $(\mu_{\rm S} = 2\Delta_\tx{ind},V_{\rm Z} = 0.8\Delta_\tx{ind})$ which lies close to the phase transition where the gap [shown in (c)] is smaller. One observes (slightly overlapping) Majorana bound states, localized at the ends of the system. The localization length of the Majorana states is smaller when the gap is larger, as expected. As can be seen in (a), for both these points a system uniformly covered by a SC would have been in the trivial phase.
\label{fig:1DL<<LAM}}
\end{figure*}

In this limit one can project out the higher harmonics and keep only the $m=n=0$ block of the Bloch Hamiltonian, Eq.~\eqref{eq:Bloch_H}. This results in the low-energy effective Hamiltonian
\begin{equation}\label{eq:H_eff}
\begin{split}
\mathcal{H}^{\rm eff}(k) =
\left(\frac{k^2}{2m^\ast} - \mu_0\right)\tau_z  + \alpha k\sigma_y\tau_z + B\sigma_x + \Delta_0 \tau_x.
\end{split}
\end{equation}
Namely, we obtain the Hamiltonian of a uniform system, where the chemical potential and induced pairing potential are given by the spatial averages of $\mu(x)$ and $\Delta(x)$, respectively. By plugging the expressions for $\mu_0$ and $\Delta_0$ from Eq.~\eqref{eq:Fourier_expan} in the topological criterion for a uniform system~\cite{PhysRevLett.105.077001,PhysRevLett.105.177002}, Eq.~\eqref{eq:uniform condition 1D}, we conclude that in the short-lattice-constant regime the system is topological whenever
\begin{equation}\label{eq:fast_mod_top_criter}
V_{\rm Z}^2>r^2\left[\left(\frac{1-r}{r}\mu_N+\mu_S\right)^2+\Delta_{\rm ind}^2\right].
\end{equation}

This result can be understood intuitively. In the short-lattice-constant regime, where the unit cell $L$ is much shorter than all other length scales in the problem (in particular the electron wavelength), the electron \emph{effectively} experiences the averages of the modulated chemical potential and induced pairing potential.

\begin{figure*}
\begin{tabular}{cc}

\includegraphics[scale=0.45]{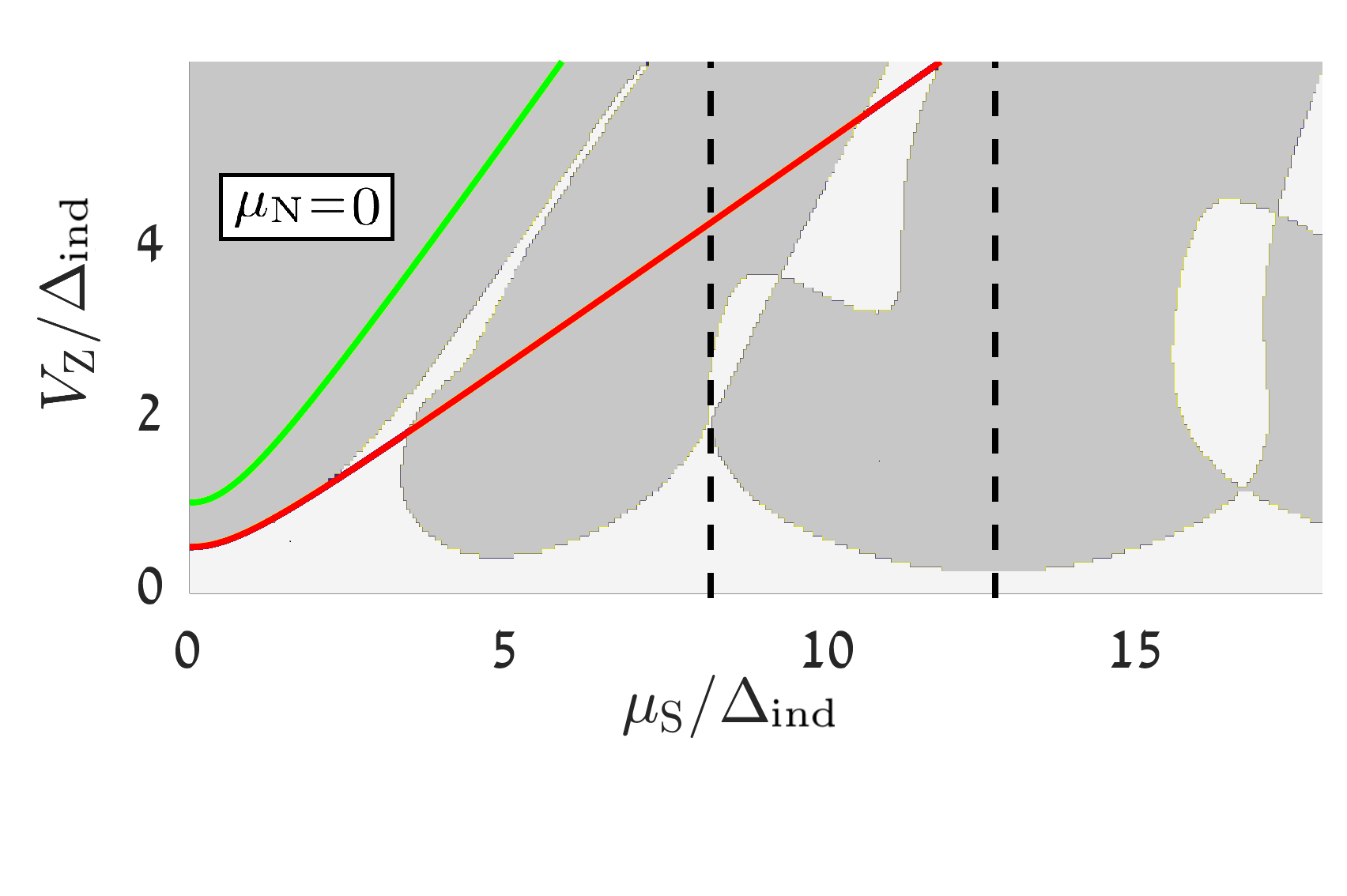}
\llap{\parbox[c]{13.3cm}{\vspace{-9.28cm}{(a)}}}
\llap{\parbox[c]{5.2cm}{\vspace{-4.1cm}{B}}}
\llap{\parbox[c]{8.8cm}{\vspace{-4.1cm}{A}}}
& \hskip 1mm

\includegraphics[scale=0.45]{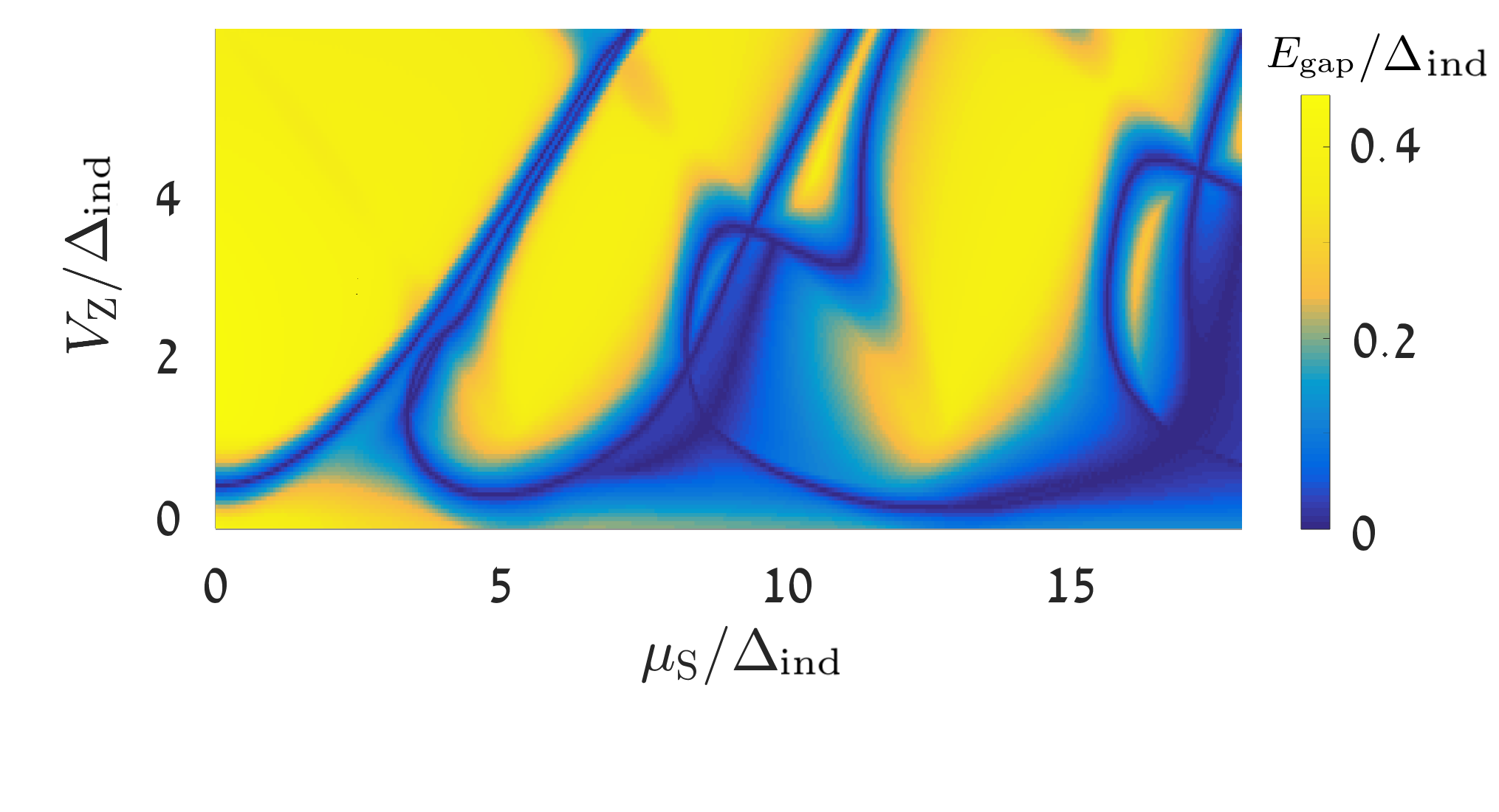}
\llap{\parbox[c]{15.9cm}{\vspace{-9.3cm}{(b)}}}

\end{tabular}
\vspace{-10mm}\caption{Numerical simulations of the system described in Eqs.~\eqref{eq:1D real space hamiltonian},~\eqref{eq:Delta_modulation} and~\eqref{eq:mu_modulation} and depicted in Fig.~\hyperref[fig:Setup]{\ref{fig:Setup}(a)}, in the regime where $E_\tx{L}=1/(m^\ast L^2)$ is of the order of other energy scales in the problem. The parameters used in the simulation are the same as those in Fig.~\ref{fig:1DL<<LAM} , except $L_{\rm S}=L/2 $ $(r=1/2)$ and the length of the superlattice unit cell is taken to be $L=830\tx{nm}\cong 6\cdot l_{\rm so}$.
(a) The topological invariant $\mathcal{Q}$ as a function of the Zeeman field, $V_{\rm Z}$, and the chemical potential in the SC regions, $\mu_{\rm S}$, calculated according to Eq.~\eqref{eq:Tight binding invariant}.
Topological ($\mathcal{Q}=-1$) regions  in parameter space are marked by dark grey and trivial ($\mathcal{Q}=1$) regions  in parameter space are marked by light grey. The additional minibands, created by the periodicity of the system, generate topological regions for chemical potentials and Zeeman fields which are outside the region obtained from the uniform approximation of Eq. \eqref{eq:fast_mod_phase_bound_muN0} (marked here in red). (b) The excitation-energy gap in units of $\Delta_\tx{ind}$.
As in the short-lattice-constant regime shown in Fig.~\hyperref[fig:1DL<<LAM]{\ref{fig:1DL<<LAM}(c)}, the gap deep in the topological regions (and for Zeeman fields which are not large compared to $E_{\rm so}$) is approximately given by $r\Delta_\tx{ind}$.\label{fig:1d_L_lso}}
\end{figure*}

The above conclusion is supported by the numerical results presented in Fig.~\ref{fig:1DL<<LAM}. We simulate the system described in Eqs.~\eqref{eq:1D real space hamiltonian},~\eqref{eq:Delta_modulation} and~\eqref{eq:mu_modulation} by discretizing it and solving the corresponding tight-binding Hamiltonian (see Appendix~\ref{sec:1d tight binding} for details). The system parameters are taken to be compatible with an InAs semiconductor wire covered (periodically) by superconducting Al. Accordingly, we take~\cite{Das2012} $E_{\rm so}=75\mu{\rm eV}$, $l_{\rm so}=1/(m^\ast\alpha)=130{\rm nm}$, and $\Delta_{\rm ind}=50\mu{\rm eV}$. We take the length of the superlattice unit cell to be $L=30\tx{nm}$, corresponding to $E_\tx{L}=2.8\tx{meV}$, and the part of the unit cell covered with a SC to be $r=L_\tx{S}/L=1/3$ [cf. Fig.~\hyperref[fig:Setup]{\ref{fig:Setup}(a)}].

The phase diagram as a function of Zeeman field $V_\tx{Z}$ and chemical potential $\mu_\tx{S}$ is presented in Fig.~\hyperref[fig:1DL<<LAM]{\ref{fig:1DL<<LAM}(a)}, for a fixed value of $\mu_\tx{N}=0$. The phase diagram is calculated by imposing periodic boundary conditions on the tight-binding Hamiltonian and using the Pfaffian topological invariant given in Eq. \eqref{eq:Kiteav invariant}. As expected for a system in the short-lattice-constant regime, where $E_\tx{L}$ is the leading energy scale, the phase diagram agrees with Eq.~\eqref{eq:fast_mod_top_criter}, which for $\mu_\tx{N}=0$ yields the phase boundary
\begin{equation}\label{eq:fast_mod_phase_bound_muN0}
V_{\rm Z}=r\sqrt{\Delta_\tx{ind}^{2}+\mu_{\rm S}^{2}},
\end{equation}
marked in red. The green line follows $V_{\rm Z}=\sqrt{\Delta_\tx{ind}^{2}+\mu_{\rm S}^{2}}$, which would have described the topological phase transition if the wire was uniformly covered by the SC. The difference between the red and green lines represents the enlargement of the topological region in parameter space, due to the reduced average chemical and induced pairing potentials in the case of the SC lattice.

As noted earlier, in practice, $\mu_{\rm S}$ is typically larger than the threshold value of $\sqrt{V_{\rm Z}^{2}-\Delta_\tx{ind}^{2}}$, below which a uniform system would be in the topological phase [see Eq.~\eqref{eq:uniform condition 1D}]. Furthermore, $\mu_{\rm S}$ is not easily controlled in the experiment, due to electrostatic screening by the SC.
In this respect, we note that the factor of $r<1$ in Eq.~\eqref{eq:fast_mod_phase_bound_muN0}
allows the system to be brought into the topological phase for weaker magnetic fields and for higher values of $\mu_{\rm S}$, as is demonstrated in Fig.~\hyperref[fig:1DL<<LAM]{\ref{fig:1DL<<LAM}(a)}.

More generally, since $\mu_\tx{S}$ cannot be easily controlled, it is crucial to be able to tune to the topological phase by controlling $\mu_\tx{N}$ for a given fixed value of $\mu_\tx{S}$. This is indeed the case, as demonstrated in Fig.~\hyperref[fig:1DL<<LAM]{\ref{fig:1DL<<LAM}(b)}, where for any value of $\mu_\tx{S}$ the topological phase can be reached by varying $\mu_\tx{N}$. In the short-lattice-constant regime considered here, this can also be inferred from Eq.~\eqref{eq:fast_mod_top_criter}. The resulting expression for the phase boundary,
\begin{equation}
\mu_\tx{N} = -\frac{r}{1-r}\mu_\tx{S}  \pm
\frac{1}{1-r}\sqrt{V_{\rm Z}^2-r^2\Delta_\tx{ind}^{2}},\label{eq:muN muS L_<<_Lambda condiiton}
\end{equation}
is marked in red, and agrees with the phase diagram in Fig.~\hyperref[fig:1DL<<LAM]{\ref{fig:1DL<<LAM}(b)}.

In Fig.~\hyperref[fig:1DL<<LAM]{\ref{fig:1DL<<LAM}(c)} we present the excitation energy gap as a function of $V_\tx{Z}$ and $\mu_\tx{S}$. The gap closes at the phase transition line, and then reopens as a function of either $V_\tx{Z}$ or $\mu_{S}$. We note that, away from the phase transition line,
the size of the gap protecting the Majorana bound states is reduced in comparison to the case of uniform SC. This is simply due to the reduction of the average pairing potential of $\Delta_0=r\Delta_\tx{ind}$, by the periodic modulation of $\Delta(x)$.

\begin{figure}
\begin{tabular}{cc}
\includegraphics[scale=0.325]{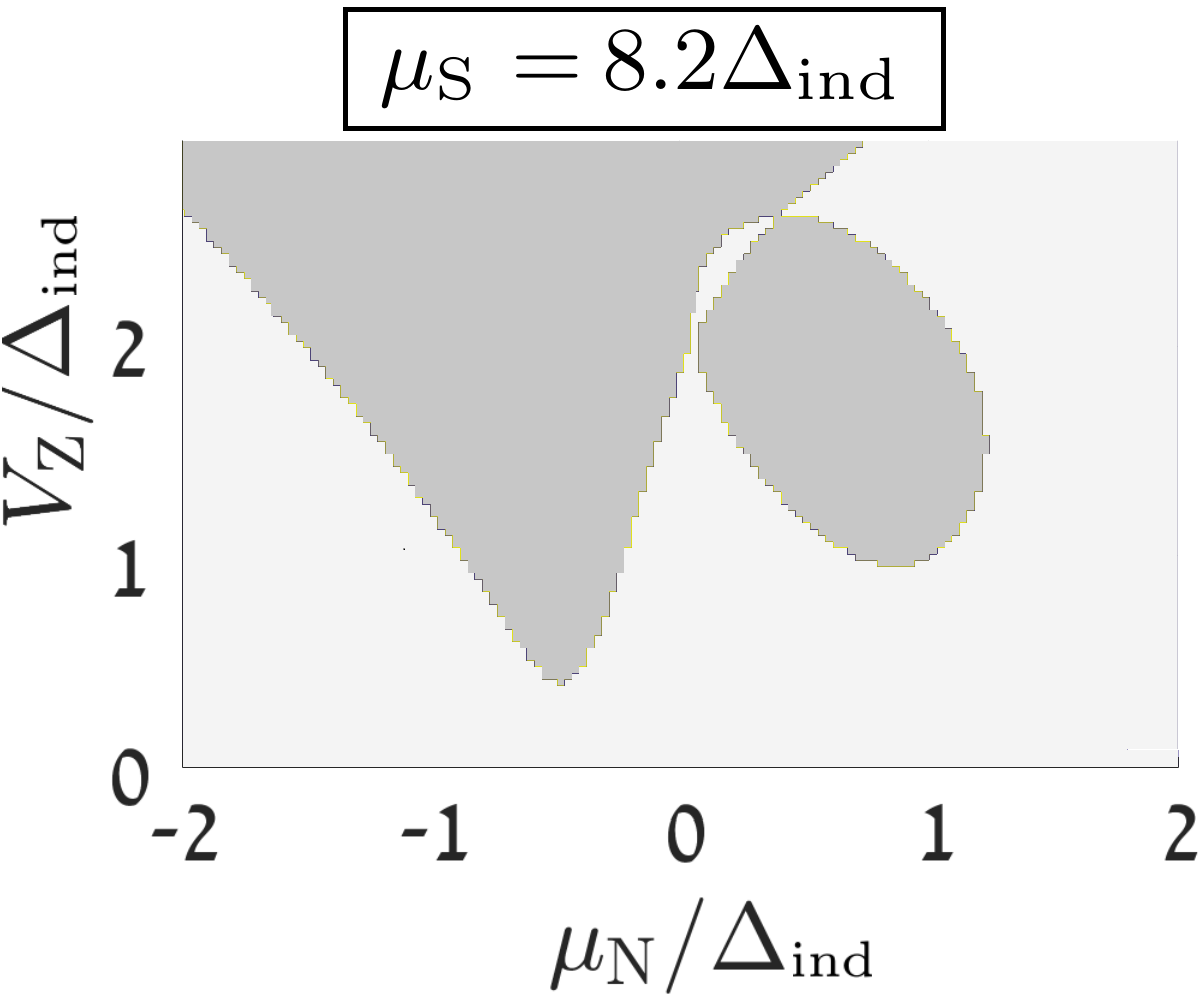}
\llap{\parbox[c]{6.53cm}{\vspace{-5.3cm}{(a)}}}

& \hskip 1mm

\includegraphics[scale=0.325]{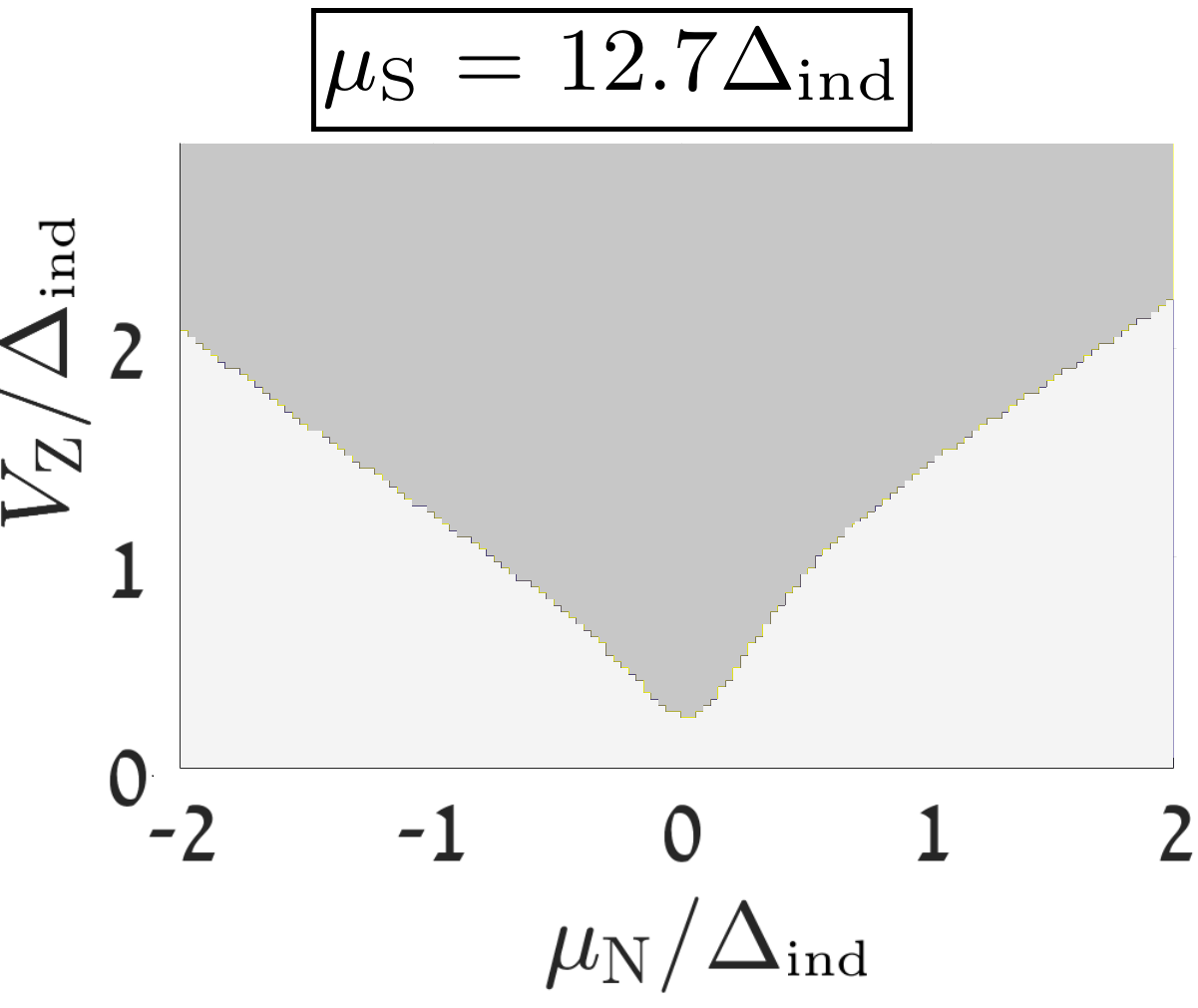}
\llap{\parbox[c]{6.5cm}{\vspace{-5.25cm}{(b)}}}

\end{tabular}
\caption{The topological invariant $\mathcal{Q}$ as a function of the experimentally accessible parameters $V_{\rm Z}$ and
 $\mu_{\rm N}$, calculated
according to Eq.~\eqref{eq:Tight binding invariant}, for $L=830\tx{nm}\cong 6\cdot l_{\rm so}$ and
$L_{\rm S}=L/2$. Topological ($\mathcal{Q}=-1$) regions in parameter space are marked by dark grey and trivial ($\mathcal{Q}=1$) regions in parameter space are marked by light grey. The chemical potential in the SC region is fixed at a value of (a) $\mu_{\rm S}=8.2\Delta_{\rm ind}$
[line A in Fig.~\hyperref[fig:1d_L_lso]{\ref{fig:1d_L_lso}(a)}] and (b)
$\mu_{\rm S}=12.7\Delta_{\rm ind}$ [line B in Fig.~\hyperref[fig:1d_L_lso]{\ref{fig:1d_L_lso}(a)}]. As can be seen in Fig.~\hyperref[fig:1d_L_lso]{\ref{fig:1d_L_lso}(a)}, at line A the system is in a region
of a trivial phase in parameter space ``in between'' regions of
a topological phase for low and more accessible values of $V_{\rm Z}$, while at line B the system can be brought to a topological phase for these values of $V_{\rm Z}$.
It can be seen that by varying the experimentally-accessible parameters
$\mu_{\rm N}$ and $V_{\rm Z}$, one can reach the topological phase even
without the ability to change $\mu_{\rm S}$, for both situations.}\label{fig :VZMUN}
\end{figure}

In Fig.~\hyperref[fig:1DL<<LAM]{\ref{fig:1DL<<LAM}(d)}, the local density of states (LDOS) at the energy of the first excited state, $E_\tx{M}=0.001\Delta_{\rm ind}$ for point $A$ and $E_\tx{M}=0.02\Delta_{\rm ind}$ for point $B$, is shown for the two points marked on the phase diagram of Fig.~\hyperref[fig:1DL<<LAM]{\ref{fig:1DL<<LAM}(a)}. The Majorana bound states are clearly visible at the ends of the wire.  Notice that both points are located between the red and the green lines in
Fig.~\hyperref[fig:1DL<<LAM]{\ref{fig:1DL<<LAM}(a)}, meaning that for these parameters a system uniformly covered by a SC would have been in the trivial phase.

\subsection{Effect of Increasing the Lattice-Constant}
\label{sec: 1D LgtrsimLambda}

We now move on to consider the more general case where the length of the unit cell, $L$, is comparable or larger than other relevant length scales. Accordingly, $E_\tx{L}=1/(m^\ast L^2)$ is no longer considered to be the leading energy scale in the problem, and the approximation leading to the effective Hamiltonian in Eq.~\eqref{eq:H_eff} does not hold. An example of such a scenario is presented in Fig.~\hyperref[fig:minibands]{\ref{fig:minibands}(b)} which shows the electronic part of the spectrum of the Bloch Hamiltonian, Eq.~\eqref{eq:Bloch_H}. The minibands, formed by the mixing of the different harmonics [labeled by $m$ and $n$ in Eq.~\eqref{eq:Bloch_H}] are now clearly visible.

As one allows for the (average) chemical potential to sweep through the bands, we expect that for small $\Delta_\tx{ind}$ the system will be in the topological phase whenever there is an odd number of pairs of Fermi points. There should therefore be a series of topological phase transitions as a function of either $\mu_\tx{N}$ or $\mu_\tx{S}$. Such a situation is indeed observed in Fig.~\ref{fig:1d_L_lso}, which presents results for the topological index and the excitation gap as a function of $V_\tx{Z}$ and $\mu_\tx{S}$, for $L=830\tx{nm}\cong 6\cdot l_{\rm so}$ and $r=L_{\rm S}/L=1/2$. The rest of the system parameters are the same as in Fig.~\hyperref[fig:1DL<<LAM]{\ref{fig:1DL<<LAM}(a)}.

Focusing first on the phase diagram in Fig.~\hyperref[fig:1d_L_lso]{\ref{fig:1d_L_lso}(a)}, the effect of the mini-bands is visible as they generate topological ($\mathcal{Q}=-1$) regions for chemical potentials, $\mu_{\rm S}$, much higher than that of the uniform case, Eq. \eqref{eq:fast_mod_phase_bound_muN0}. This creates an interesting situation in which the use of a super lattice
seemingly grants access to the topological phase for a wider range of $\mu_{\rm S}$, compared with the uniform case.

It is important to note that the observed higher minibands are not the transverse sub-bands that become relevant when the cross section of the wire become larger. Namely, they appear even if only one transverse channel is taken into consideration (as is the case here) and, as explained, are the result of the super-lattice structure.

Since the parameter $\mu_\tx{S}$ is not easily controlled in the experiment, and is mostly determined by the properties of the SC-semiconductor interface, it is important to consider the sensitivity of the phase diagram to $\mu_\tx{\rm _S}$.
Let us concentrate, for example, on lines A and B marked in Fig.~\hyperref[fig:1d_L_lso]{\ref{fig:1d_L_lso}(a)}. For line A $(\mu_{\rm S}=8.2\Delta_{0})$, the system is
in between
lobes of topological regions, and a large Zeeman field is required in order for the system to enter the topological phase. For line B $(\mu_{\rm S}=12.7\Delta_{0})$ on the other hand, the system
is driven into the topological phase for lower, more accessible, values of the Zeeman field.

In Figs.~\hyperref[fig :VZMUN]{\ref{fig :VZMUN}(a)} and ~\hyperref[fig :VZMUN]{\ref{fig :VZMUN}(b)} we study the phase diagram as a function of the
chemical potential in the normal regions, $\mu_\tx{N}$ (which can be controlled by a gate potential), and the Zeeman field, $V_\tx{Z}$,
for the two values of $\mu_\tx{S}$ corresponding to line A and line B, respectively. Importantly, Fig.~\hyperref[fig :VZMUN]{\ref{fig :VZMUN}(a)} shows that even for the more ``problematic'' value of $\mu_{\rm S}$ [line A in Fig.~\hyperref[fig:1d_L_lso]{\ref{fig:1d_L_lso}(a)}], one can still reach the topological phase by tuning the experimentally-accessible parameters, $\mu_\tx{N}$ and $V_\tx{Z}$.

\subsection{\label{sec:SOC in SC}Spin-Orbit Coupling in the Superconducting Region Only}

\begin{figure}
\begin{tabular}{cc}
\includegraphics[scale=0.26]{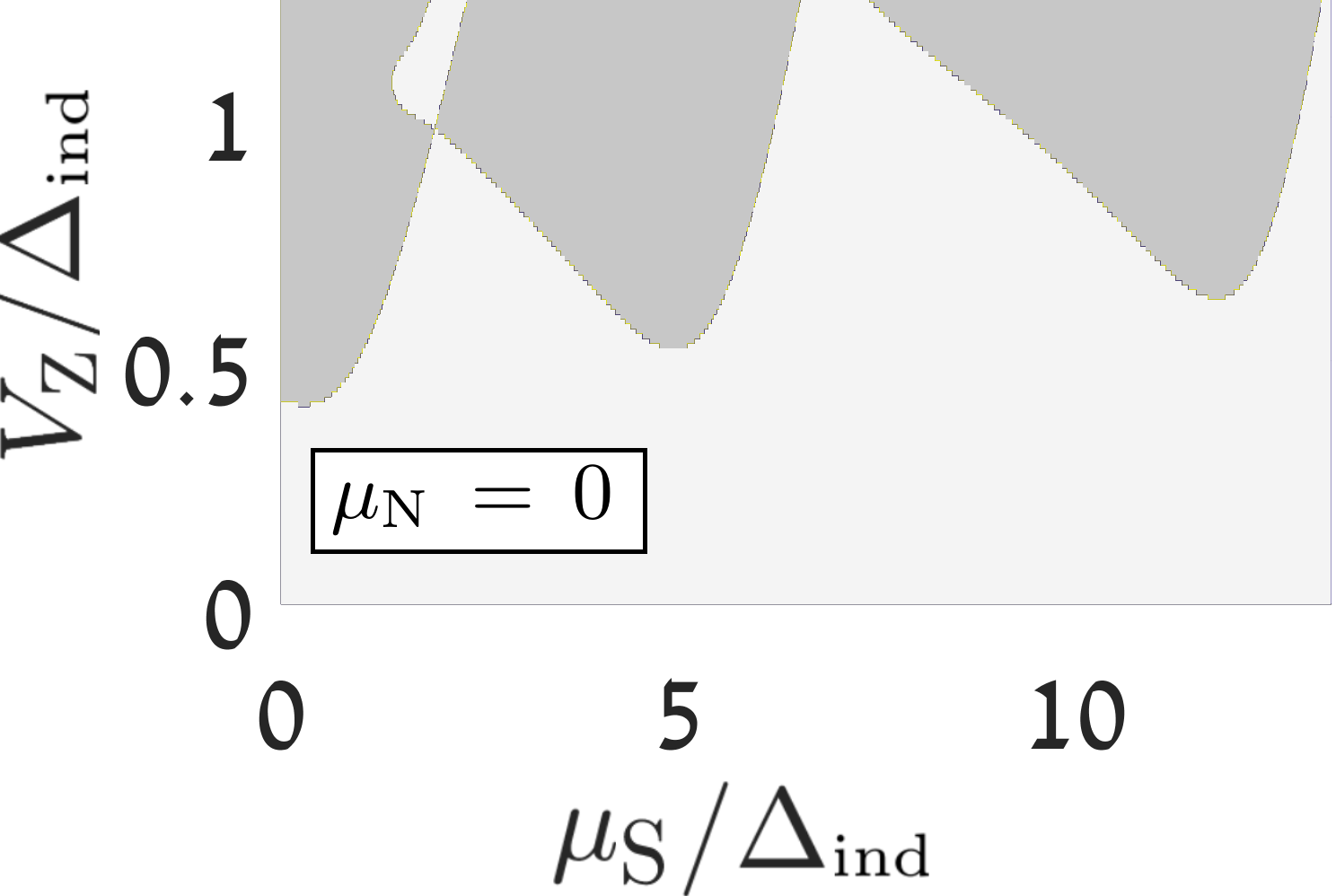}
\llap{\parbox[c]{6.25cm}{\vspace{-5cm}{(a)}}}

& \hskip 1mm

\includegraphics[scale=0.26]{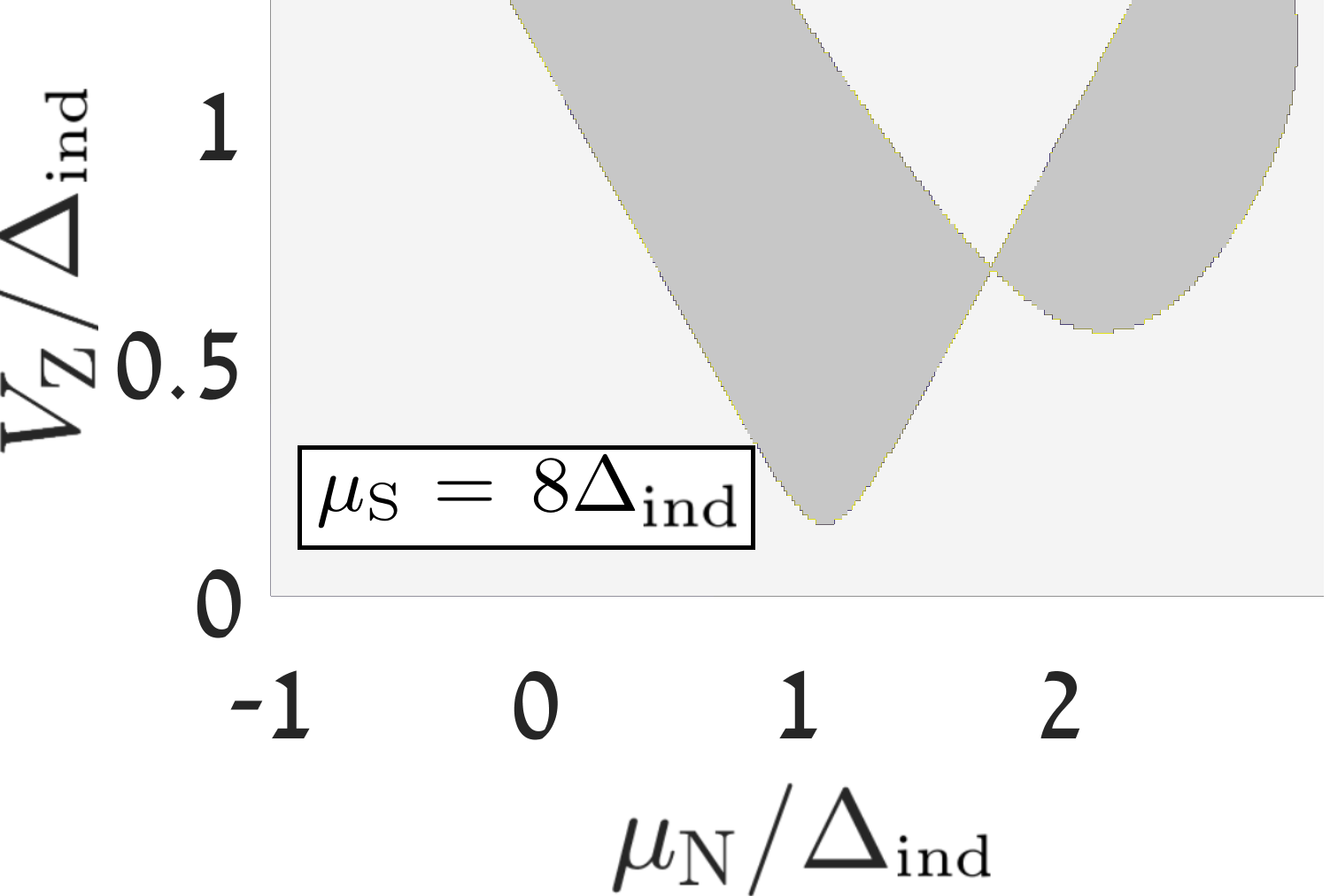}
\llap{\parbox[c]{6.25cm}{\vspace{-5cm}{(b)}}}

\end{tabular}
\caption{\label{fig:SOC in SC}The topological invariant $\mathcal{Q}$ for a one dimensional SC lattice in
which the SOC exists only in regions covered by a SC. The length of the lattice unit cell is $L=830\tx{nm}\cong 6\cdot l_{\rm so}$,
and the part covered by the SC is $r=L_{\rm S}/L=2$. Topological ($\mathcal{Q}=-1$) regions in parameter space are marked by dark grey and trivial ($\mathcal{Q}=1$) regions in parameter space are marked by light grey. (a) The topological invariant
as a function of $\mu_{\rm S}$ and $V_{\rm Z}$ for fixed $\mu_{\rm N}=0$ and (b) The topological invariant
as a function of $\mu_{\rm N}$ and $V_{\rm Z}$ for fixed $\mu_{\rm S}=8\Delta_{\rm ind}$. As can be seen in (a), for this value of $\mu_{\rm S}$ the system is in a region of a trivial phase in parameter space ``in between'' regions of a topological phase for low and more accessible values of $V_{\rm Z}$. It can be seen that by varying the experimentally accessible parameters $\mu_{\rm N}$ and $V_{\rm Z}$, one can reach the topological phase even without the ability to change $\mu_{\rm S}$.}
\end{figure}

We end this section by considering the case in which the SOC exists only in the regions covered by a SC. This can happen if the SC has strong SOC, but not the semiconductor. In this scenario, the SC induces both superconductivity and SOC.
We demonstrate that in this case the system can be successfully brought to the topological regime by varying the experimentally-accessible parameters $\mu_{\rm N}$ and $V_{\rm Z}$, even when $\mu_{\rm S}$ is in the trivial regime for a uniform system.

This scenario can resemble the case of a one-dimensional system composed of light chemical elements, such as a Carbon nanotube which has no substantial SOC, covered by a SC lattice composed of a heavy SC such as NbN. Furthermore, this example suggests that, for the two-dimensional case (to be examined below), one could use
a heavy SC in proximity to a 2DEG with weak SOC, such as GaAs.

As before, we model the system by discretizing it and constructing a tight-binding model (see Appendix~\ref{sec:App tight binding}),
where now the SOC exists only in the sites belonging to the regions covered by a SC. In the remaining sites which are not in proximity to a SC, we take the Rashba SOC strength to be $\alpha=0$. In the short-lattice-constant regime, where $L$ is the shortest length scale, it was demonstrated in Eq.~\eqref{eq:H_eff} that one obtains an effectively uniform Hamiltonian, as the electrons experience 'smeared' potentials. Thus, one expects in this case a phase diagram according to Eq.~\eqref{eq:fast_mod_top_criter}.

We therefore concentrate on the less trivial case where $L$ is of the order of other length scales in the problem (such as the Fermi wavelength, coherence length etc.). Fig.~\hyperref[fig:SOC in SC]{\ref{fig:SOC in SC}(a)} shows the phase diagram obtained for $L=830\tx{nm}\cong 6\cdot l_{\rm so}$, where the spin-orbit-coupling length in the SC region was taken to be as in the previous section. Similar lobes of topological regions in parameter space as those analyzed in Sec.~\ref{sec: 1D LgtrsimLambda} appear at high values of $\mu_{\rm S}.$ Indeed, we see that even beyond the short-lattice-constant regime we obtain a topological phase for a SC lattice that has SOC only in regions covered by a SC. The SC lattice scheme also allows us to tune into the topological phase using the gate potential in the normal regions, as is demonstrated in Fig.~\hyperref[fig:SOC in SC]{\ref{fig:SOC in SC}(b)}.

\section{a two-dimensional superconducting lattice \label{sec:2D_phase}}

\subsection{The Model\label{sec:2D_model}}

\begin{figure*}
\begin{tabular}{cc}

\includegraphics[scale=0.205]{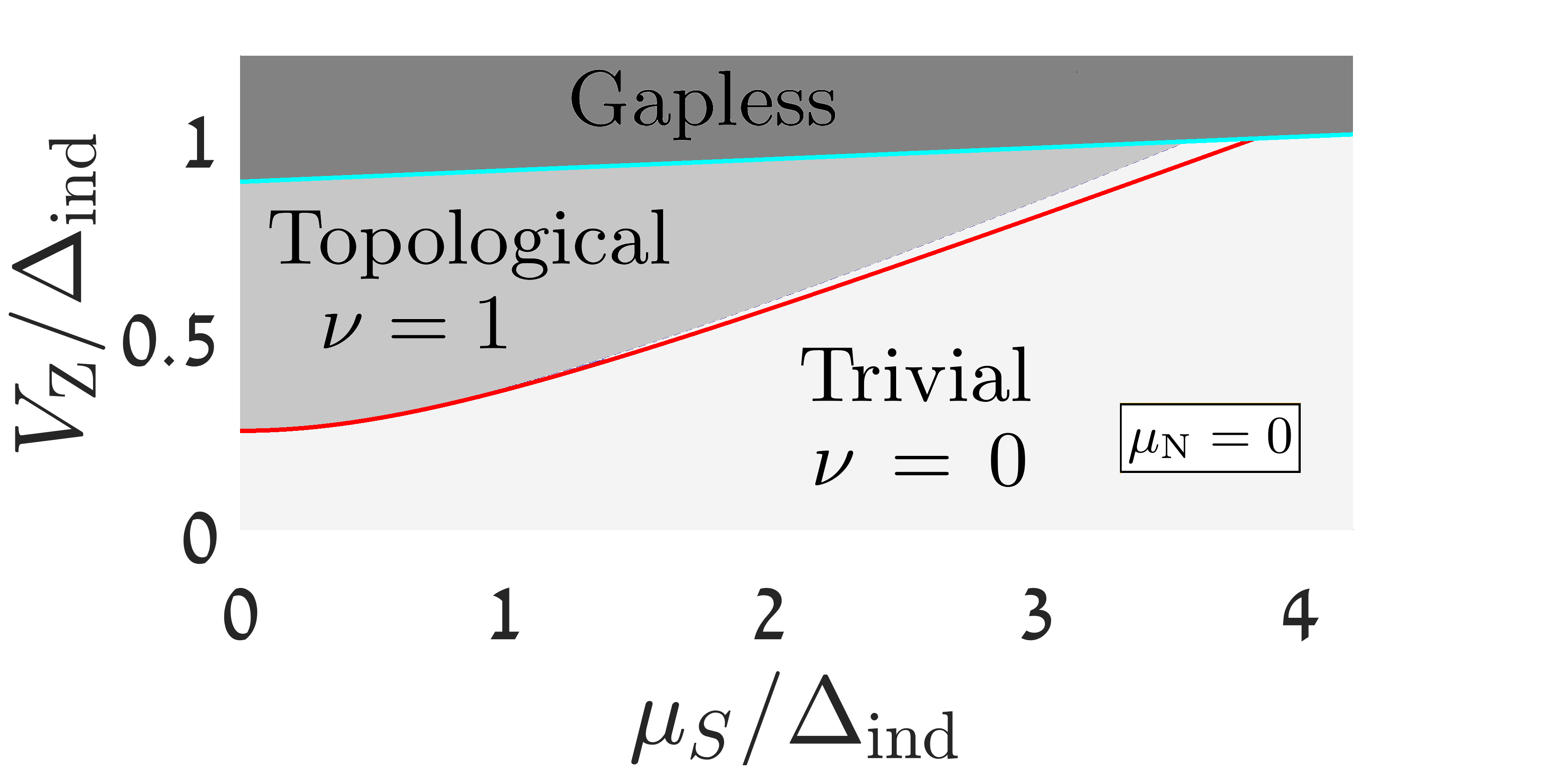}
\llap{\parbox[c]{12.8cm}{\vspace{-6.55cm}{\color{white}(a)}}}

& \hskip 1mm

\includegraphics[scale=0.205]{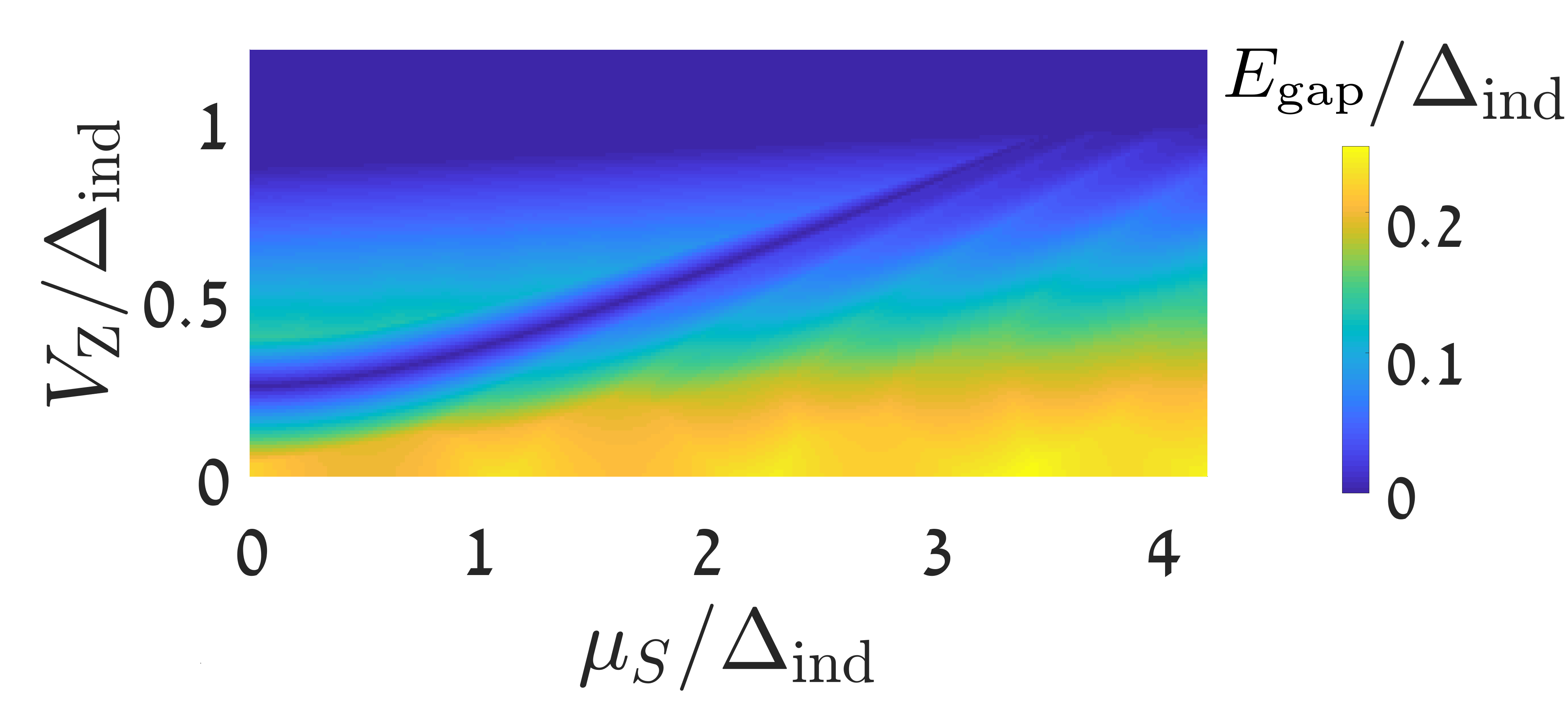}
\llap{\parbox[c]{14.15cm}{\vspace{-6.5cm}{\color{white}(b)}}}

 \\

\includegraphics[scale=0.205]{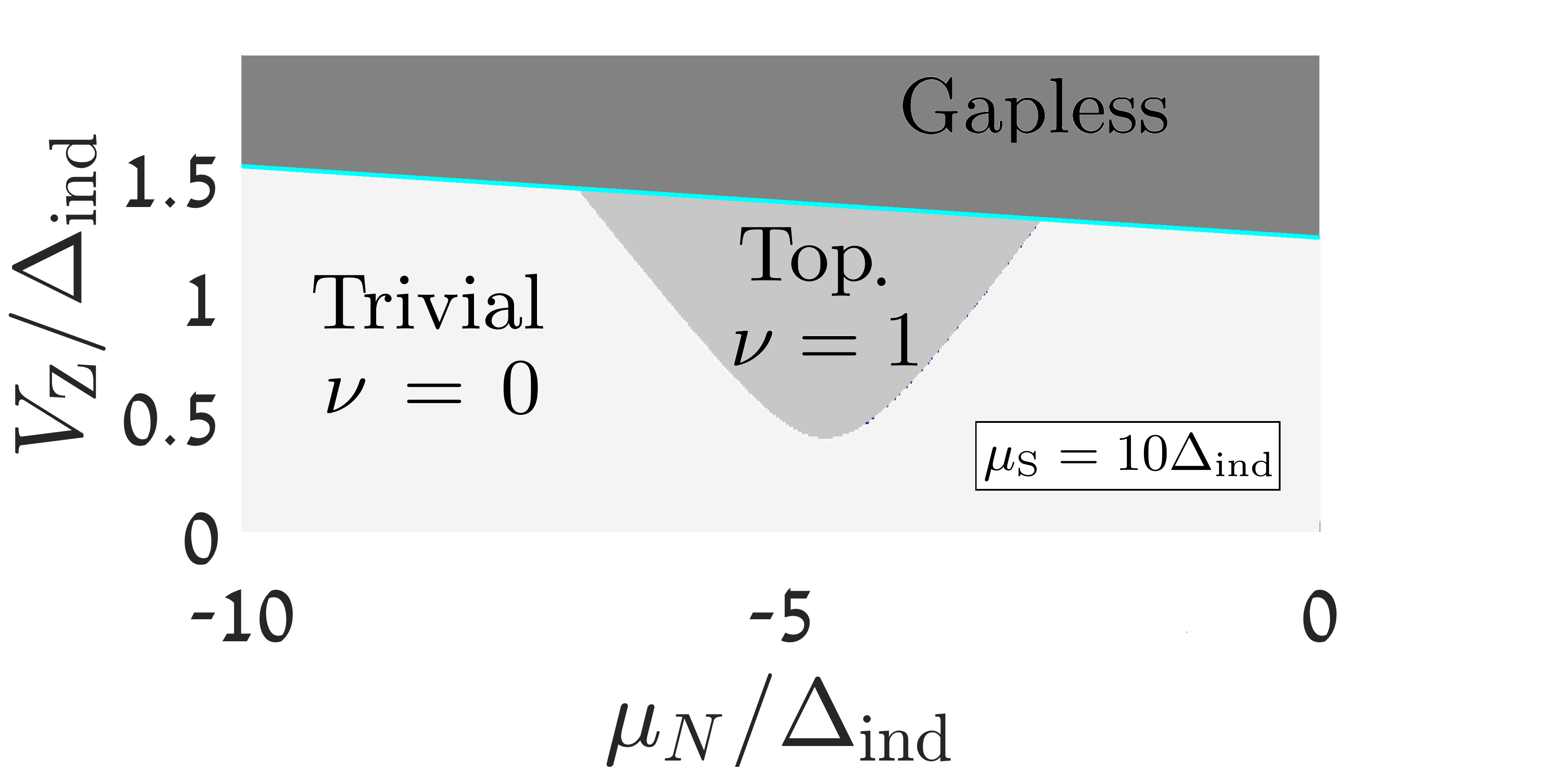}
\llap{\parbox[c]{12.8cm}{\vspace{-6.55cm}{\color{white}(c)}}}

& \hskip 1mm

\includegraphics[scale=0.205]{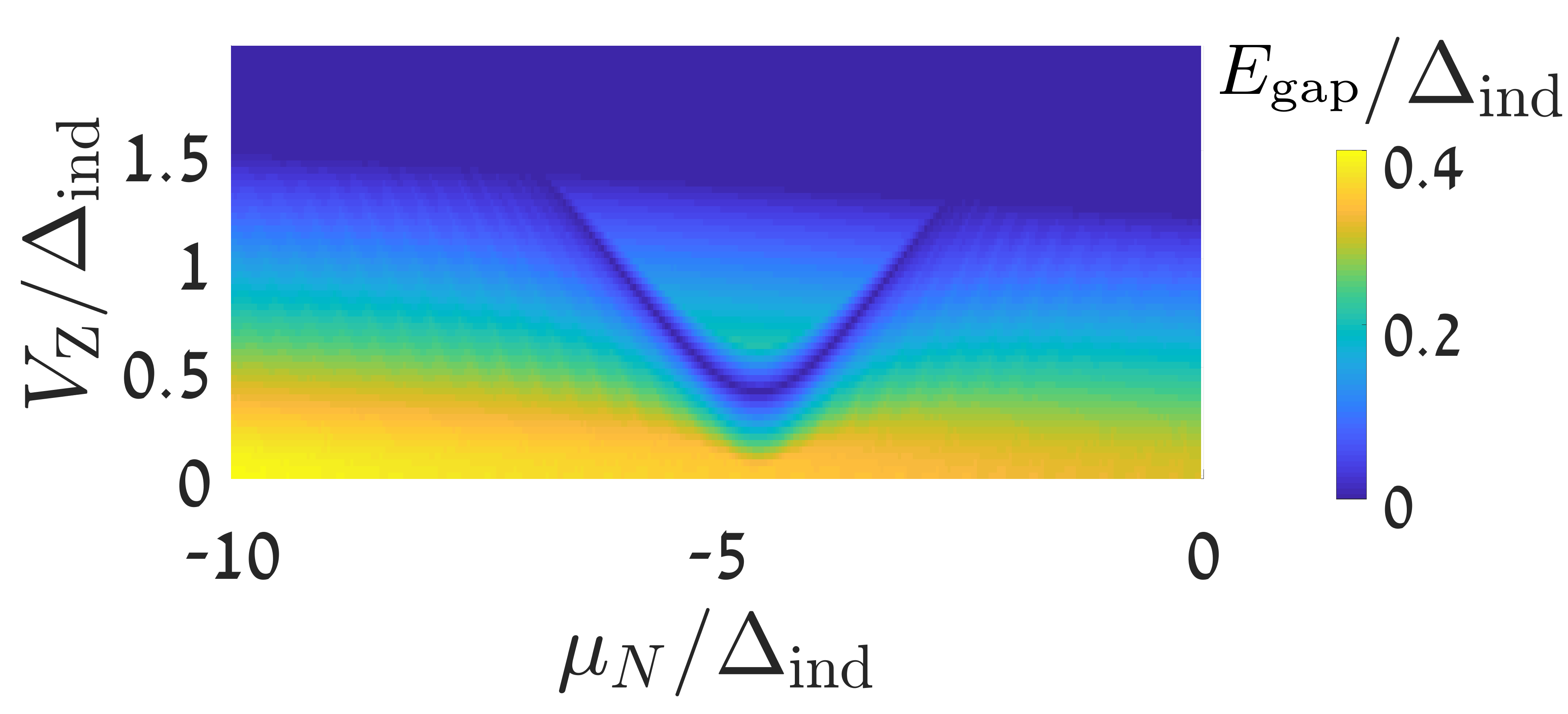}
\llap{\parbox[c]{14.1cm}{\vspace{-6.5cm}{\color{white}(d)}}}

\end{tabular}
\caption{Numerical simulations of the system covered by a two-dimensional SC lattice such as depicted in Fig.~\hyperref[fig:Setup]{\ref{fig:Setup}(b)}, in the regime where $E_L=1/(m^\ast L_{x}^2)=1/(m^\ast L_{y}^2)$ is of the order of other energy scales in the problem.  The system, that is described by Eqs.~\eqref{eq: 2D real space hamiltonian} and~\eqref{eq:mu_delta_2d_profile}, has Rashba and Dresselhaus SOC. For a similar system that is uniformly covered by a SC, an in-plain magnetic field can drive the system into a topological $p_x+ip_y$ phase, as suggested by Ref.~[\onlinecite{PhysRevB.81.125318}]. We demonstrate that this phase can be achieved by using a SC lattice which overcomes the challenge of gating the system in the presence of a uniform SC. The SOC strength ratio between the Rashba and Dresselhaus terms is taken to be:~\cite{Dressel} $\alpha/\beta=0.3$, and the corresponding SOC lengths are taken to be $l^\tx{R}_{\rm so}=1/(m^\ast\alpha)={130}{\rm nm}$ and $l^\tx{D}_{\rm so}=1/(m^\ast\beta)={39}{\rm nm}$. The induced pairing potential is $\Delta_{\rm ind}={50}\mu{\rm eV}$. The SC lattice constants are $L_x=L_y={100}{\rm nm}$, and the part of the unit cell covered with a SC is $r=L_{\tx{S}_x}L_{\tx{S}_y}/(L_xL_y)={0.25}$.
(a) The phase diagram as a function of the in-plane Zeeman field, $V_{\rm Z}$, and the chemical potential in the SC regions, $\mu_{\rm S}$ when the chemical potential in the non-SC regions, $\mu_{\rm N}$ is tuned to $0$. The Chern number is calculated according to Eq. \eqref{eq:Chern_number}. Trivial ($\nu=0$) regions in parameter space are marked by light grey, topological ($\nu=1$) regions in parameter space are marked by darker grey and regions in which the system is in the gapless phase are marked by the darkest grey. For higher in-plane Zeeman fields, a gapless phase is obtained due to a tilt in the spectrum (see Fig.~\ref{fig:2d_majorana}). The phase transition to the gapless phase is marked by the cyan line. For lower in-plane Zeeman fields, the topological phase is obtained in a region similar to the one in the short unit cell approximation of Eq.~\eqref{eq:uniform condition 2d} that is marked in red, with $\Delta_0=r\Delta_\tx{ind}$ and $\mu_0=r\mu_\tx{S}$. Deviations from the short unit cell approximation are apparent near the transition to the gapless phase.
(b) The excitation-energy gap in units of $\Delta_\tx{ind}$, which
closes exactly at the phase transition line. (c)  The phase diagram for a scenario in which the chemical potential in the SC regions, which is dictated by the materials and is hard to control due to screening effects, has a high value of $\mu_{\rm S}=10\Delta_{\rm ind}$. The phase diagram is presented as a function of the in-plane Zeeman field, $V_{\rm Z}$, and the chemical potential in the non-SC regions, $\mu_{\rm N}$, which are experimentally accessible parameters. The results demonstrate that even when the interface between the semiconductor and the SC dictates a high chemical potential in areas covered by the SC, the topological phase can be obtained in the SC lattice geometry by only controlling the chemical potential in the non-SC regions via an external gate. (d) A corresponding excitation-energy gap in units of $\Delta_\tx{ind}$.}\label{fig :2DGap}
\end{figure*}

We now move on to study the two-dimensional (2d) case. We are interested in using a SC lattice geometry in order to realize the topological $p_x+ip_y$ phase. We consider a 2d electron gas (2DEG), with strong SOC, in proximity to a SC with a periodic structure, as depicted in Fig.~\hyperref[fig:Setup]{\ref{fig:Setup}(b)}. As in the 1d case, the SC lattice geometry is meant to enable control over the chemical potential in order to overcome electrostatic screening by the SC.

For the case of a 2DEG having a Rashba SOC and \emph{uniformly} covered by a SC, it was shown~\cite{Sau2010} that the system is driven into the topological phase by applying an out-of-plane Zeeman field. This can in principle be accomplished by coupling the 2DEG to a ferromagnetic insulator which then induces a Zeeman coupling. Achieving the same out-of-plane Zeeman coupling by applying a magnetic field is accompanied by orbital effects which create vortexes and modify the state of the SC itself.

It was suggested, however, that in the presence of Dresselhaus SOC, an \emph{in-plane} Zeeman coupling can also drive the system into the topological phase~\cite{PhysRevB.81.125318}. Being in the in-plane direction, the Zeeman coupling can now readily be achieved by the application of a magnetic field, with negligible orbital effects. This scheme has further advantages, as it lacks the technical challenge of coupling the 2DEG to a ferromagnet and it enables tuning into the topological phase by controlling the strength of the magnetic field. A 2DEG with an appreciable Dresselhaus SOC (in addition to Rashba SOC) can be achieved by growing the semiconductor in the (110) direction~\cite{PhysRevB.81.125318}.

In this section we adopt this setup of a semiconductor with both Rashba and Dresselhaus type SOC, and study the case where it is \emph{periodically} covered by a conventional SC. In the presence of an in-plane magnetic field, the Hamiltonian describing the system is given by
\begin{equation}
\begin{split}
&H=\int \tx{d}^2{\bf r} \Psi^\dagger({\bf r}) \mathcal{H}({\bf r}) \Psi({\bf r})~;\\
&\mathcal{H}({\bf r}) = \left[-\frac{\nabla^2}{2m^\ast}-\mu({\bf r})+i\alpha(\sigma_y\partial_{x}-\sigma_x\partial_{y})-i\beta\sigma_z\partial_{x}\right]\tau_z\\
& \hskip 7.5mm + V_{\rm Z}\sigma_y + \Delta({\bf r})\tau_x,
\end{split}\label{eq: 2D real space hamiltonian}
\end{equation}
with $\Psi^\dagger({\bf r}) = [\psi^\dagger_\uparrow({\bf r}),\psi^\dagger_\downarrow({\bf r}),\psi_\downarrow({\bf r}),-\psi_\uparrow({\bf r})]$, where $m^\ast$ is the electron effective mass, $\mu({\bf r})$ is the chemical potential, $\alpha$ is the Rashba SOC coefficient, $\beta$ is the Dresselhaus SOC coefficient, $V_{\rm Z}$ is the Zeeman field, and $\Delta({\bf r})$ is the proximity-induced pairing potential.

For a uniform chemical potential, $\mu\left({\bf r}\right)=\mu_{\rm 0}$, and a uniform pairing potential, $\Delta\left({\bf r}\right)=\Delta_{0}$, the Hamiltonian in Eq. \eqref{eq: 2D real space hamiltonian} coincide with the one analyzed in Ref. [\onlinecite{PhysRevB.81.125318}]. In this case the system is in the topological $p_x+ip_y$ phase for~\cite{PhysRevB.81.125318}
\begin{equation}
|\mu_{\rm 0}|<\sqrt{V_{\rm Z}^{2}-\Delta_{0}^{2}}\equiv\mu_{\rm c},\label{eq:uniform condition 2d}
\end{equation}
as long as the bulk spectrum is gapped.

The pairing potential and the chemical potential considered in this work are described by
\begin{equation}
\begin{split}
\Delta({\bf r}) = \Delta_{\rm ind} &\hskip -2mm \sum_{n_x,n_y\in\mathbb{Z}} \hskip -2mm  {\rm rect}\left(\frac{x-n_xL_x}{L_{{\rm S}_x}}\right){\rm rect}\left(\frac{y-n_yL_y}{L_{{\rm S}_y}}\right),\\
\mu({\bf r}) = \mu_\tx{N} +& (\mu_\tx{S}-\mu_\tx{N}) \times \\
\times &\hskip -2mm \sum_{n_x,n_y\in\mathbb{Z}} \hskip -2mm  {\rm rect}\left(\frac{x-n_xL_x}{L_{{\rm S}_x}}\right){\rm rect}\left(\frac{y-n_yL_y}{L_{{\rm S}_y}}\right).
\end{split}\label{eq:mu_delta_2d_profile}
\end{equation}
where $L_x\times L_y$ are the dimensions of the unit cell, of which $L_{{\rm S}_x}\times L_{{\rm S}_y}$ is covered by a superconductor [see also Fig.~\hyperref[fig:Setup]{\ref{fig:Setup}(b)}]. As in the 1d case, $\mu_{\rm S}$ ($\mu_{\rm N}$) is the chemical potential in the parts (not) covered by the SC. Similar to the 1d case, the transition between $\mu_N$ and $\mu_S$ may be smoother, an effect which
we have not included in the present calculation but is not expected to qualitatively affect the results.

The Hamiltonian in Eq.~\eqref{eq: 2D real space hamiltonian} obeys a particle-hole symmetry, $\tau_y\sigma_y\mathcal{H}^\ast\tau_y\sigma_y=-\mathcal{H}^\ast$, putting it in symmetry-class D~\cite{PhysRevB.55.1142} with a $\mathbb{Z}$ topological invariant~\cite{kitaev2009periodic,PhysRevB.78.195125}, known as the Chern number. As in the 1d case of Sec.~\ref{sec:1D-superconducting-lattice}, we solve the model by first discretizing the Hamiltonian (see details in Appendix~\ref{sec: 2d TB}). Upon diagonalizing the resulting tight-binding Hamiltonian, we can obtain the spectrum, and the Chern number, as explained in Eq.~\eqref{eq:Chern_number} in Appendix~\ref{app:sc_lattice_k}.

\begin{figure}

\includegraphics[scale=0.23]{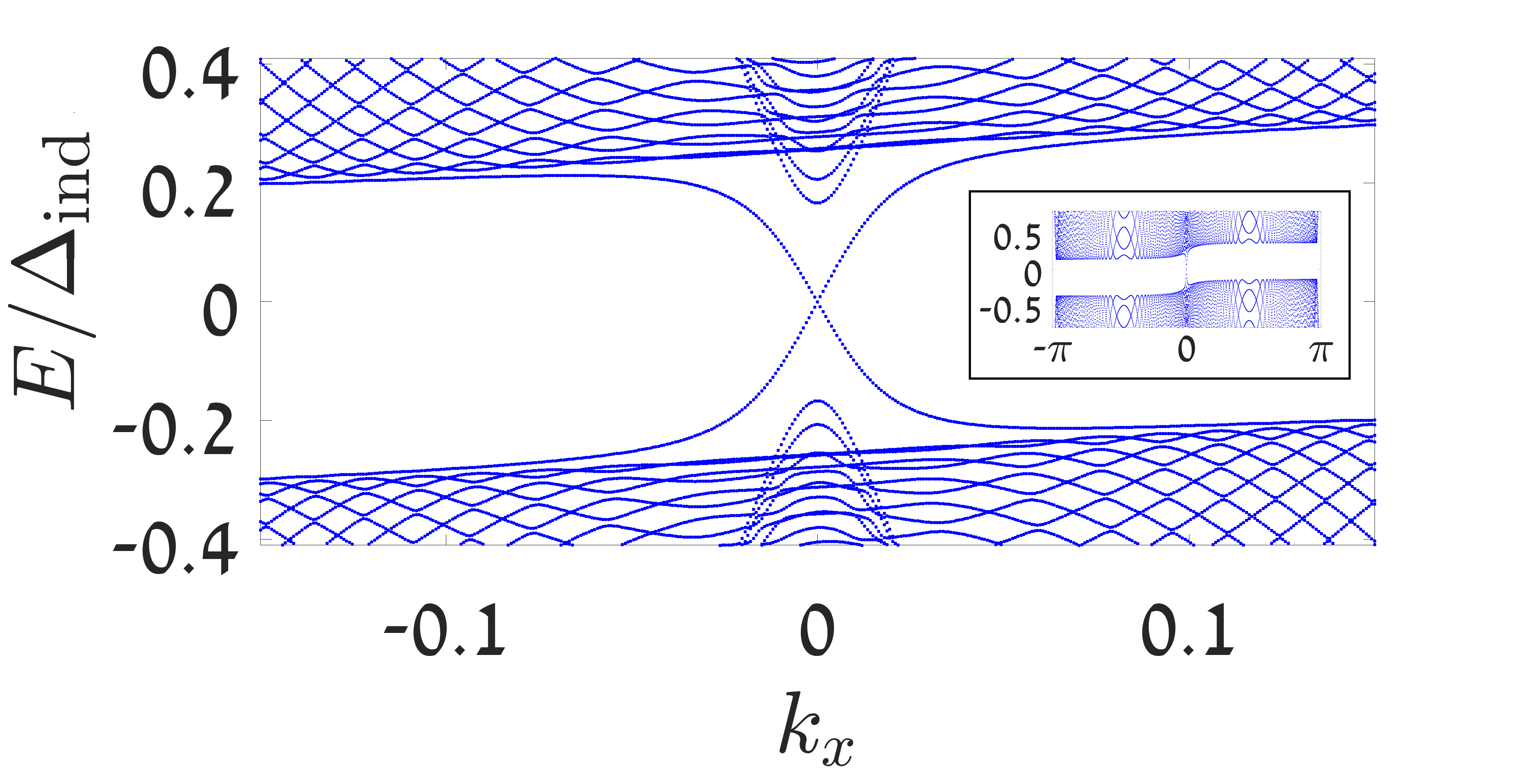}

\caption{The spectrum as a function of $k_{x}$ for a system with the same parameters as in Fig.~\ref{fig :2DGap},
in a cylinder geometry with periodic boundary conditions along the $x$ direction and open boundary conditions along the $y$ direction. The spectrum is calculated for $\mu_\tx{S}=0$ and $V_\tx{Z}=20\mu\tx{eV}=0.4\Delta_{\rm ind}$, a point that would have been well in the trivial phase for a system uniformly covered by a SC (see Eq.~\eqref{eq:uniform condition 2d}). The height of the cylinder was taken here to be ${12{\rm \mu m}}$.
The gapless and chiral Majorana edge modes are visible at the center
of the spectrum, counter propagating at the two ends of the cylinder. The region showing the propagating Majorana modes is enlarged, the entire spectrum is shown in the inset. The apparent tilt in the spectrum is due to the in-plane Zeeman field. For sufficiently large values of the field, the gap closes at $k_x,k_y \neq 0$ and the system enters a gapless phase.}\label{fig:2d_majorana}
\end{figure}

As in the 1d case, in limit where the lattice constants, $L_x$, $L_y$, are much shorter than all other relevant length scales, i.e. in the limit where $E_L=1/(m^\ast L_{x,y}^2)$ is the leading energy scale, we expect the behavior of the system to be determined by the average values of the chemical and paring potentials, $\mu_0$ and $\Delta_0$. The phase boundary between the trivial and the topological phase is then described by Eq.~\eqref{eq:uniform condition 2d}.

\subsection{Simulation Results}

We investigate the more realistic case of $L_x, L_y \sim l_{\rm so}$ numerically in Fig.~\ref{fig :2DGap}. Figs.~\hyperref[fig :2DGap]{\ref{fig :2DGap}(a)} and~\hyperref[fig :2DGap]{\ref{fig :2DGap}(b)} show the phase diagram and the energy gap as a function of the chemical potential in the SC region, $\mu_\tx{S}$, and the Zeeman field, $V_\tx{Z}$, where the chemical potential in the normal regions is fixed at $\mu_\tx{N}=0$. The SOC strength ratio between the Rashba and Dresselhaus terms is chosen as:~\cite{Dressel} $\alpha/\beta=0.3$, and the corresponding SOC lengths are taken to be $l^\tx{R}_{\rm so}=1/(m^\ast\alpha)={130}{\rm nm}$ and $l^\tx{D}_{\rm so}=1/(m^\ast\beta)={39}{\rm nm}$.  The induced pairing potential is $\Delta_{\rm ind}={50}\mu{\rm eV}$. The SC lattice constants are $L_x=L_y={100}{\rm nm}$, and the part of the unit cell covered with a SC is $r=L_{\tx{S}_x}L_{\tx{S}_y}/(L_xL_y)={0.25}$.

As can be seen in Fig.~\hyperref[fig :2DGap]{\ref{fig :2DGap}(a)}, for relatively low values of the in-plane Zeeman field, a phase boundary follows Eq.~\eqref{eq:uniform condition 2d} quite well. The line of Eq.~\eqref{eq:uniform condition 2d} is marked in red, with $\Delta_0=r\Delta_\tx{ind}$ and $\mu_0=r\mu_\tx{S}$. A gapless phase emerges at higher values of the in-plane Zeeman field, as the gap closes at $k_x,k_y \neq 0$ due to a tilt in the spectrum that is caused by the in-plane field [cf. Fig.~\ref{fig:2d_majorana}]. The deviation of the actual phase transition line (that occurs at the gap closure) from the form of Eq.~\eqref{eq:uniform condition 2d} (red line), that is expected in this $L_x, L_y \sim l_{\rm so}$ regime, is apparent near the transition to the gapless phase. Indeed, in the 1d case the deviation from  the uniform case behavior was apparent at relatively high values of the Zeeman field [cf. Fig.~\hyperref[fig:1d_L_lso]{\ref{fig:1d_L_lso}(a)}], which are not accessible in this setup.

Fig.~\hyperref[fig :2DGap]{\ref{fig :2DGap}(b)} shows that the phase transition occurs for sufficiently low values of the Zeeman field such that a topological region with a gap of $E_{\rm gap}\cong 0.15\Delta_{\rm ind}$ is obtained. As in the 1d case, the topological region in parameter space is expanded to more accessible parameter values, namely lower Zeeman fields and higher values of $\mu_{\rm S}$, due to the reduction of the effective pairing potential by a factor of $r$.

Having demonstrated an ability to tune into the topological phase for low values of the chemical potential under the SC, $\mu_{\rm S}$, we turn to ask what happens when this parameter, which is dictated by the materials and is hard to control due to screening effects, has a high value. Figures.~\hyperref[fig :2DGap]{\ref{fig :2DGap}(c)} and~\hyperref[fig :2DGap]{\ref{fig :2DGap}(d)}, show that even when $\mu_{\rm S}=10\Delta_{\rm ind}$, which is well in the trivial zone for
$\mu_{\rm N}=0$ [see Fig.~\hyperref[fig :2DGap]{\ref{fig :2DGap}(a)}], the topological phase can be obtained by varying the chemical potential in the normal regions, which is an experimentally accessible knob. Besides the chemical potentials, Figs.~\hyperref[fig :2DGap]{\ref{fig :2DGap}(c)} and~\hyperref[fig :2DGap]{\ref{fig :2DGap}(d)} have the same simulation parameters as Figs.~\hyperref[fig :2DGap]{\ref{fig :2DGap}(a)} and~\hyperref[fig :2DGap]{\ref{fig :2DGap}(b)}.

Finally, we can examine the chiral Majorana edge modes that are predicted to exist in the topological $p_x+ip_y$ phase. To this end we consider the same system in a cylinder geometry with periodic boundary conditions along the $x$ direction and open boundary conditions along the $y$ direction, for $\mu_\tx{S}=0$ and $V_\tx{Z}=20\mu\tx{eV}$. For a system uniformly covered by a SC, this point would have been well in the trivial phase, as the phase boundary line receives a minimum for $V_{\rm Z}= \Delta_{\rm ind}=50\mu\tx{eV}$ in that case [see Eq.~\eqref{eq:uniform condition 2d}]. As can be seen in Fig.~\ref{fig:2d_majorana}, there are two counter-propagating modes crossing the Fermi energy in an otherwise gapped system. These are the two Majorana modes, one at each end of the cylinder.

\section{Discussion}~\label{sec:discussion}

In this work, we have studied the possibility of realizing a topological superconductor by using SC lattice [see Fig.~\ref{fig:Setup}], thereby overcoming the problem of controlling the density in the vicinity of a SC. While the chemical potential in the regions covered by the SC, $\mu_\tx{S}$, is difficult to control due to screening, the chemical potential in the uncovered regions, $\mu_\tx{N}$, is an experimentally-accessible parameter. We have demonstrated that the topological phase can be reached even in the extreme case of absolute screening where is $\mu_\tx{S}$ is fixed. Namely, that gating in the normal regions, which controls $\mu_\tx{N}$, is sufficient in order to tune into the topological phase.

We have analyzed different regimes of the superlattice constant, $L$. For $L$ much smaller than all other relevant length scales, the system is effectively described by a uniform Hamiltonian with the average values of the chemical and induced pairing potentials [see Fig.~\ref{fig:1DL<<LAM}].
When $L$ becomes of the order of other length-scales (such as the SOC length, $l_\tx{so}$, the coherence length, $\xi$, and the Fermi wavelength $\lambda$, for which a reasonable estimate in proximity-coupled semiconductors would be  $l_\tx{so}, \lambda, \xi \sim 100\tx{nm}$),
high-energy mini-bands create additional topological regions
at higher values of $\mu_{\rm S}$ [see Fig. \ref{fig:1d_L_lso}]. This creates a situation in which larger and more realistic values of the SC lattice constant, $L$, bring forth an additional benefit in the form of more accessible topological regions in parameter space.

In two dimensions, we investigated a practical realization scheme in which an in-plane magnetic field is applied in a semiconductor that has Rashba and Dresselhaus SOC and is proximity coupled to a SC lattice. Fig.~\ref{fig :2DGap} shows that the topological $p_x+ip_y$ phase can indeed be attained in this setup, and the SC lattice can provide a solution to the gating problem in two-dimensional setups as well.

\begin{figure}
\begin{tabular}{cc}
\includegraphics[scale=0.205]{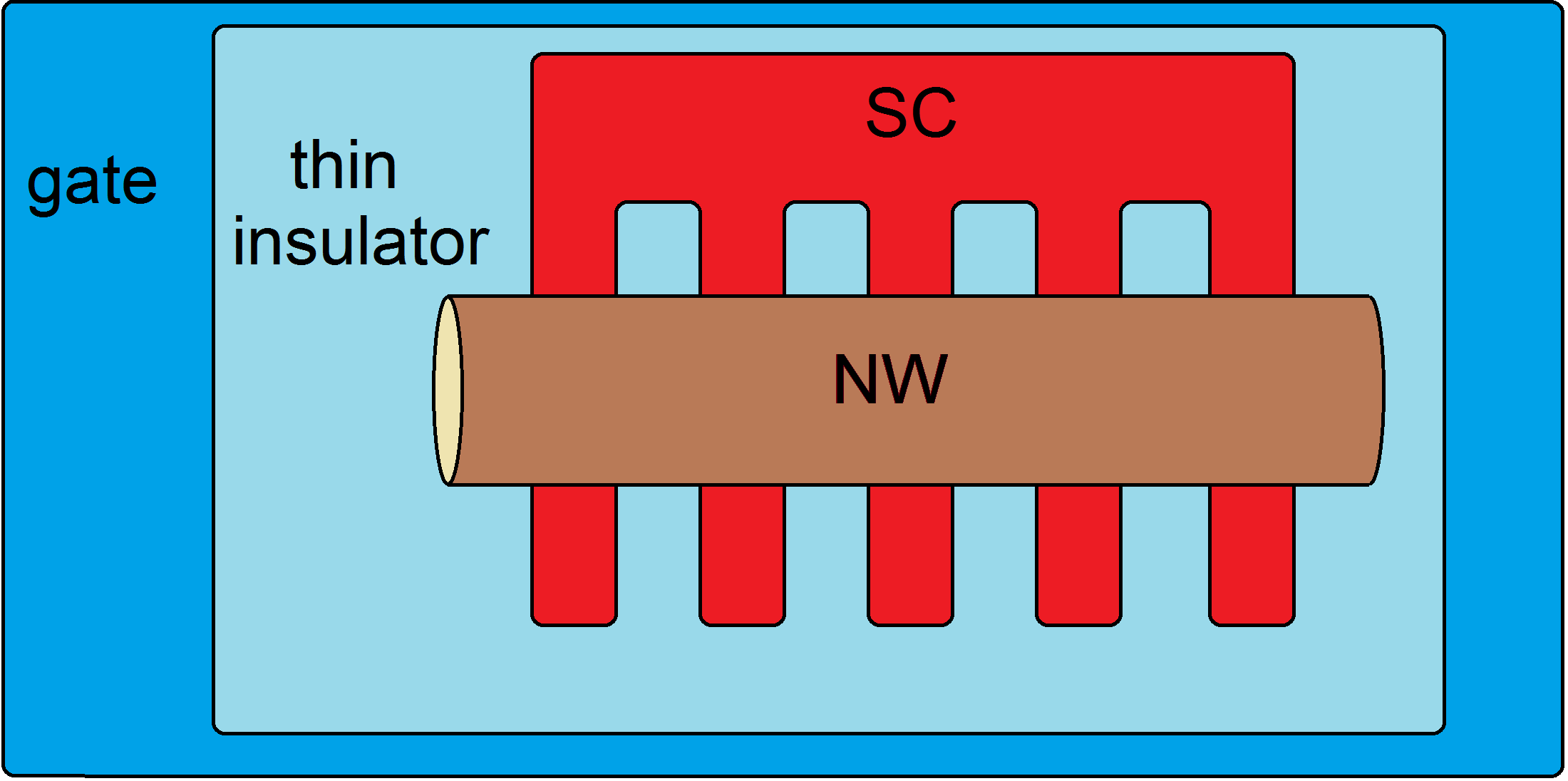}
\llap{\parbox[c]{9.22cm}{\vspace{-4.12cm}{(a)}}}

& \hskip 1mm

\includegraphics[scale=0.40]{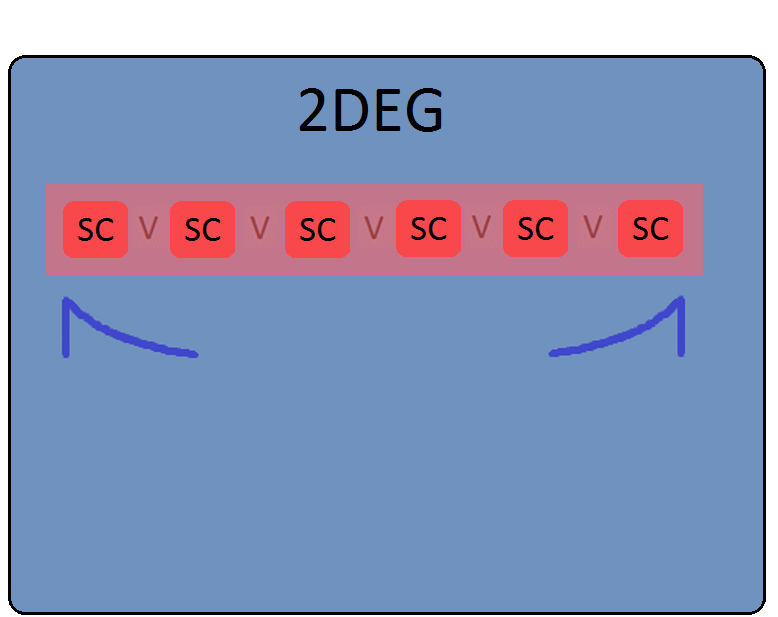}
\llap{\parbox[c]{6.2cm}{\vspace{-4.12cm}{(b)}}}

\end{tabular}
\caption{Suggested SC lattice based realization schemes for
the one-dimensional topological SC phase discussed in Sec. \ref{sec:1D-superconducting-lattice}.
(a) A nano-wire is placed \textit{above} a SC in the shape of a comb, therefore its
top can be measured by an STM without the obstruction of the SC. Gating can then be performed as in the model analyzed in Sec.
\ref{sec:1D_model}, which was shown to be in the topological regime
even if the chemical potential under the SC is completely unaffected
by the gate. (b) A 1d channel embedded in a 2DEG.
The light blue surface represents the two-dimensional substrate and
the dark blue lines represent the decaying envelope of the Majorana wave functions located at the edges of the effective ``wire''. Gating the normal regions such that $\mu_{\rm N}$ assumes the value required for the topological phase (cf. Fig. \ref{fig :VZMUN}) results in an effective
topological ``wire''. If we imagine that the SC islands are connected from above and shorted, similarly to the comb geometry presented in (a), then a measurement of the LDOS on one of the ``wire'' ends through a tunneling barrier is possible without charging the islands.\label{fig:experimental}}
\end{figure}
One can imagine realizing the one-dimensional SC phase described in Sec. \ref{sec:1D-superconducting-lattice} by periodically covering a one-dimensional nanowire with
an s-wave SC from atop. This would be in accordance with the majority
of current realization schemes, which cover the nanowire with a uniform
layer of a SC. We suggest an alternative version, depicted in Fig.~\hyperref[fig:experimental]{\ref{fig:experimental}(a)},
where the wire is placed on top of a SC in the shape of a comb. This
apparatus allows one to measure the LDOS of the Majorana states,
such as the one we have shown in Fig.~\hyperref[fig:1DL<<LAM]{\ref{fig:1DL<<LAM}(d)}, using
a scanning tunneling microscope (STM). The STM tip could access the
wire from atop, without the obstruction of the SC.

In this setup, the screening by the SC, in the regions it does exist in the comb geometry, would be substantial, rendering the gate's affect on the chemical potential negligible in these regions. Indeed,
without the use of the SC lattice geometry, it makes more sense to cover the
nanowire with the SC from atop in order to reduce the screening effects.
The demonstrated ability to tune into the topological phase by controlling the chemical potential only in the non-SC regions, allows the SC lattice based scheme to have one side of the wire accessible for an STM measurement.

Aa an alternative to the nano-wire based realization, one can attempt to use a one-dimensional channel which
resides within a two-dimensional system such as a 2DEG.
These can be made in a very precise fashion and are relatively clean, hence we expect that disorder will be weak in such a two-dimensional system. As long as the disorder energy scale, set by the inverse of the mean free time, is smaller than the gap in the topological system, we expect that the topological phase will not be destroyed. We relegate a more quantitative study of disordered effects for future investigations.
In Fig.~\hyperref[fig:experimental]{\ref{fig:experimental}(b)} we see an illustration of such
a one-dimensional SC lattice residing within a two-dimensional sample. The 2DEG can be tuned to a chemical potential that lies in one of the trivial regions and a localized gate can tune the chemical potential along the one-dimensional channel to one of the topological regions. This can be performed similarly to Ref.  [\onlinecite{suominen2017scalable}].

This work mainly focuses on systems in which the SOC originated in a heavy semiconductors.  However, as discussed in Sec.~\ref{sec:SOC in SC}, similar considerations may apply when, instead, a heavy SC with strong SOC is used, in 1d and in 2d. One can envision Graphene or other exfoliated 2d materials covered periodically by a heavy SC such as Pb or NbN. This may enable tuning the chemical potential to a value needed to establish a topological phase, similar to the technique demonstrated throughout this paper.

Charging energy of the deposited superconducting squares [see Fig.~\hyperref[fig:Setup]{\ref{fig:Setup}(b)}] may affect the properties of the system in an interesting and profound manner. In many experimental configurations, the coupling between the SC squares, due to transmission through the 2DEG or through additional superconducting bridges, is larger than the charging energy. In these cases, the charging energy effects may be neglected, as was done in this work.

We expect that the inclusion of electron-electron interaction, for example, in the form of charging energy on the SC squares, may lead to rich and exciting physics. Dualities of setups which include MBSs, electron-electron interactions, and interesting spin liquid phases are already discussed (see for example ref.[~\onlinecite{barkeshli2015physical}]). A large charging energy on a square containing four MBS, reduces the ground-state degeneracy from four to two, forming an effective spin $1/2$ state and thus establishing a link between spin models and MBSs with interactions. Theoretically, the Toric code phase, which is dual to the Abelian phase of Kitaev's spins honeycomb model, can be stabilized in superconducting islands containing four MBSs \cite{Landau2016}. The B phase, i.e., the non – Abelian phase of the honeycomb model, was proposed to be realized in a different MBS setup with interactions \cite{barkeshli2015physical}.  In the future, it will be interesting and challenging to investigate the conditions that lead to stabilizations of these spin-liquid phases and possible novel interacting phases in this playground of superconducting lattices.

\section*{Acknowledgments}

The research was supported by the Deutsche Forschungsgemeinschaft (CRC 183), the Israel Science Foundation (ISF), the Binational Science Foundation (BSF), and the European Research Council under the European Community’s Seventh Framework Program (FP7/2007- 2013)/ERC Grant agreement No. 340210.

We would like to thank C. M. Marcus for suggesting a 1d super-lattice geometry and to  acknowledge enlightening discussions with  Gil Refael, Jason Alicea and Yuval Baum.

\appendix

\section{tight binding\label{sec:App tight binding}}

The tight-binding model allows us to numerically evaluate the topological
invariants, as well as the LDOS of the zero energy excitations for
a general form of the spatial distribution of the SC lattice. This is obtained
by discretizing the Hamiltonian under study on a lattice, where the lattice spacing
is chosen such that it is smaller than the relevant physical parameters
of the system, e.g. the Fermi wavelength, the SOC length etc. The tight-binding lattice unit cell is not to be confused
with the above discussed super-lattice unit cell -- their exact relation is presented below.

\subsection{One Dimensional Tight-Binding }\label{sec:1d tight binding}

A lattice parameter $a$ is introduced, which defines
the tight-binding lattice points $x=na$ , $n=1,...,N.$ $N$ is the number of sites,
and $Na$ represents the physical length of the entire system.
The tight-binding Hamiltonian that was used to model the spatial profile of $\Delta$ and $\mu$ as given in Eqs. \eqref{eq:Delta_modulation} and \eqref{eq:mu_modulation}, is given by:
\begin{equation}
\begin{split}
H_{\rm tb}=&\underset{n,s,s'}{\sum}\left\{c\dag_{n;s}\eps_{n;ss'}c_{n,s'}+\left[c\dag_{n;s}t_{ss'}c_{n+1,s'}+{\rm H.c.}\right]\right\}\\
+&\underset{n}{\sum}\left[\Delta_{n}c\dag_{n;\uparrow}c\dag_{n;\downarrow}+{\rm H.c.}\right]\\
\end{split}\label{eq:tight binding Hamiltonian}
\end{equation}
where
\begin{equation}
\begin{split}
&\eps_{n;ss'}=\left(2t-\mu_n\right)\delta_{ss'}+V_{\rm Z}\sigma_{ss'}^{z}\\
&t_{ss'}=-t\delta_{ss'}-iu\sigma_{ss'}^{y}\\
&\mu_n = \mu_N+(\mu_S-\mu_N)f_n\\
&\Delta_{n}=\Delta_{\rm ind}f_{n}\\
&f_{n}=\sum_{j}{\rm rect}\left(\frac{na-j\cdot L}{Ls}\right),
\end{split}
\end{equation}
and where $t=1/2ma^{2},$ $u=\alpha/2a$. The term $t_{ss'}$
represents the hopping amplitude between the sites of the wire, with
 $-t\delta_{ss'}$ originating from the kinetic term, and
$-iu\sigma_{ss'}^{y}$ originating from the Rashba SOC term. Notice
that the periodic nature is preserved, as each unit cell now composed
of $L/a$ sites of which $L_{\rm S}/a$ are in proximity
to a SC (we take $L$ and $L_{\rm S}$ to be integer multiples of $a$).

We can relate the tight-binding parameters, $t$ and $u$, to the physical
parameters, $E_{\rm so}$ and $l_{\rm so}$, by:
\begin{equation}
E_{\rm so}=\frac{u^{2}}{t}\hskip 3mm ; \hskip 3mm l_{\rm so}=a\frac{t}{u}
\label{eq:SOC TB}
\end{equation}

Given physical values for $E_{\rm so},L$ and either $m^\ast$ or $l_{\rm so}$
we can compute all the tight-binding parameters. In order for the
tight-binding model to accurately describe its continuous counterpart,
we require the bandwidth $4t$ to be larger than all other energy
scales in the system: $4t\gg E_{\rm so},\Delta_{0},V_{\rm Z},\mu_{\rm 0}$. After
choosing $t$, other parameters such as $a,N,u$ can be computed using
the above equations.

\subsection{Two Dimensional Tight-Binding }\label{sec: 2d TB}

The extension of the tight-binding procedure to two dimensions is
simple. The lattice constants $a_{x}$ and $a_{y}$ define the lattice points $x=na_{x}$
and $y=ma_{y}$. The indices $n$ and $m$ count the sites between
$1$ and $N_{x},N_{y}$, in the $\hat{x}\text{ and }\hat{y}$ directions
respectively. Thus, $N_{x}a_{x}$ and $N_{y}a_{y}$ represent the
physical lengths of the system in these respective directions. Defining the tight-binding Dresselhaus SOC term $v=\beta/2a$, the Hamiltonian in Eq. \eqref{eq: 2D real space hamiltonian}, with the
spatial profile of $\Delta$ and $\mu$ as given in Eq.~\eqref{eq:mu_delta_2d_profile},
is mapped to:
\begin{equation}
\begin{split}
H_{\rm tb}=& \underset{n,m,s,s'}{\sum}\left\{c\dag_{n,m;s}\eps_{nm;ss'}c_{n,m;s'}\right.\\
+&\left[c\dag_{n,m;s}t_{ss'}^{x}c_{n+1,m;s'}+{\rm H.c.}\right]\\
+&\left.\left[c\dag_{n,m;s}t_{ss'}^{y}c_{n,m+1;s'}+{\rm H.c.}\right]\right\}\\
+&\underset{n,m}{\sum}\left[\Delta_{n,m}c\dag_{n,m;\uparrow}c\dag_{n,m;\downarrow}+{\rm H.c.}\right],
\end{split}\label{eq: 2D tight binding Hamiltonian}
\end{equation}
where
\begin{equation}
\begin{split}
&\eps_{nm;ss'}=\left(2t-\mu_{nm}\right)\delta_{ss'}+V_{\rm Z}\sigma_{ss'}^{z}\\
&t_{ss'}^{x}=-t\delta_{ss'}-iu\sigma_{ss'}^{y}+iv\sigma_{ss'}^{z}\\
&t_{ss'}^{y}=-t\delta_{ss'}+iu\sigma_{ss'}^{x}\\
&\mu_{nm} = \mu_N+(\mu_S-\mu_N)f_{nm}\\
&\Delta_{nm}=\Delta_{\rm ind}f_{nm}\\
&f_{nm}=\underset{j,l}{\sum}{\rm rect}\left(\frac{na_{x}-j\cdot L_{x}}{L_{{\rm S} x}}\right)\cdot {\rm rect}\left(\frac{ma_{y}-l\cdot L_{y}}{L_{{\rm S} y}}\right)
\end{split}
\end{equation}

\subsection{Super-Lattice Momentum Space }\label{app:sc_lattice_k}

\begin{figure}
\bigskip{}
\subfloat[]{\includegraphics[scale=0.28]{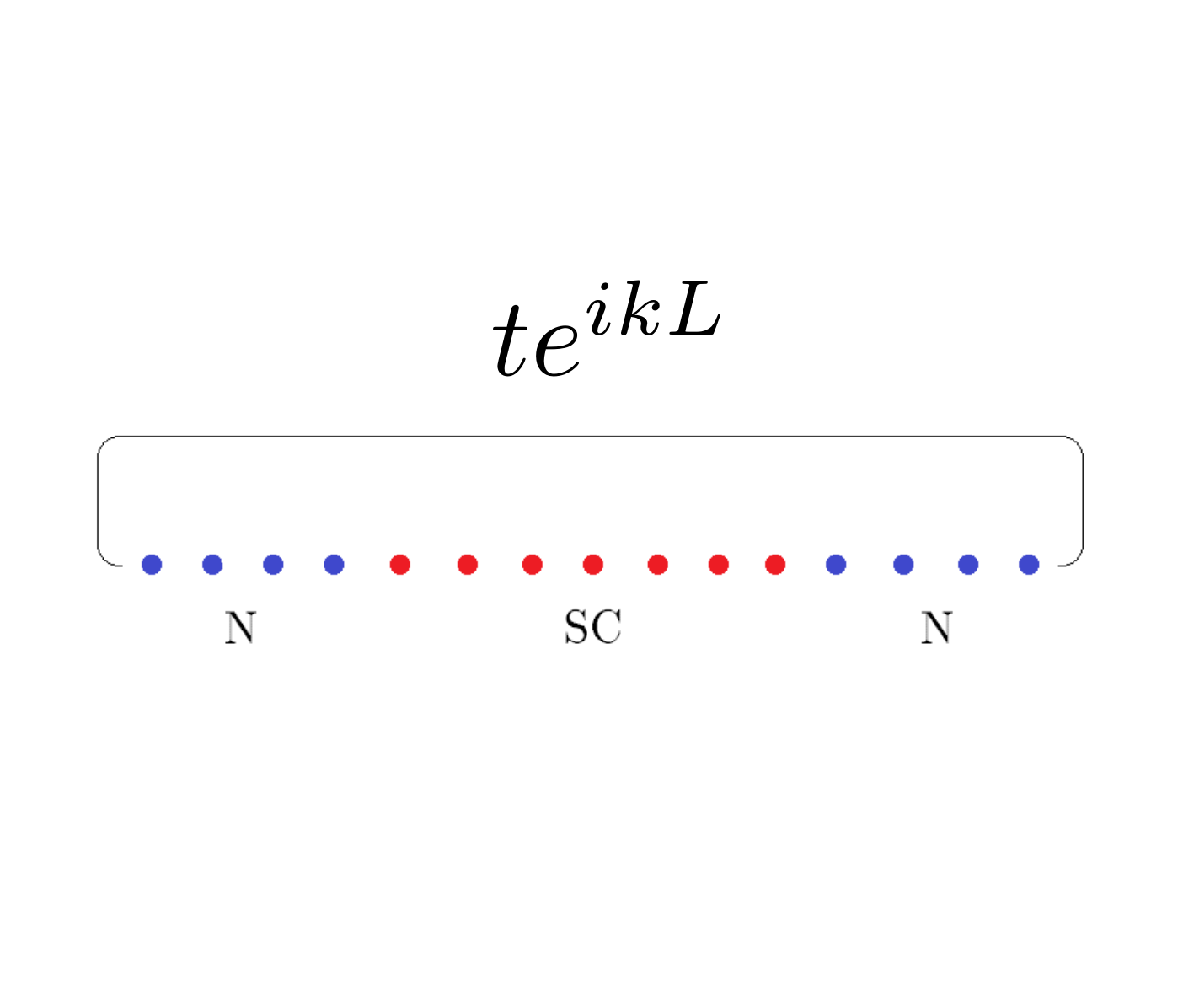}

}\hspace*{\fill}\subfloat[]{\includegraphics[scale=0.28]{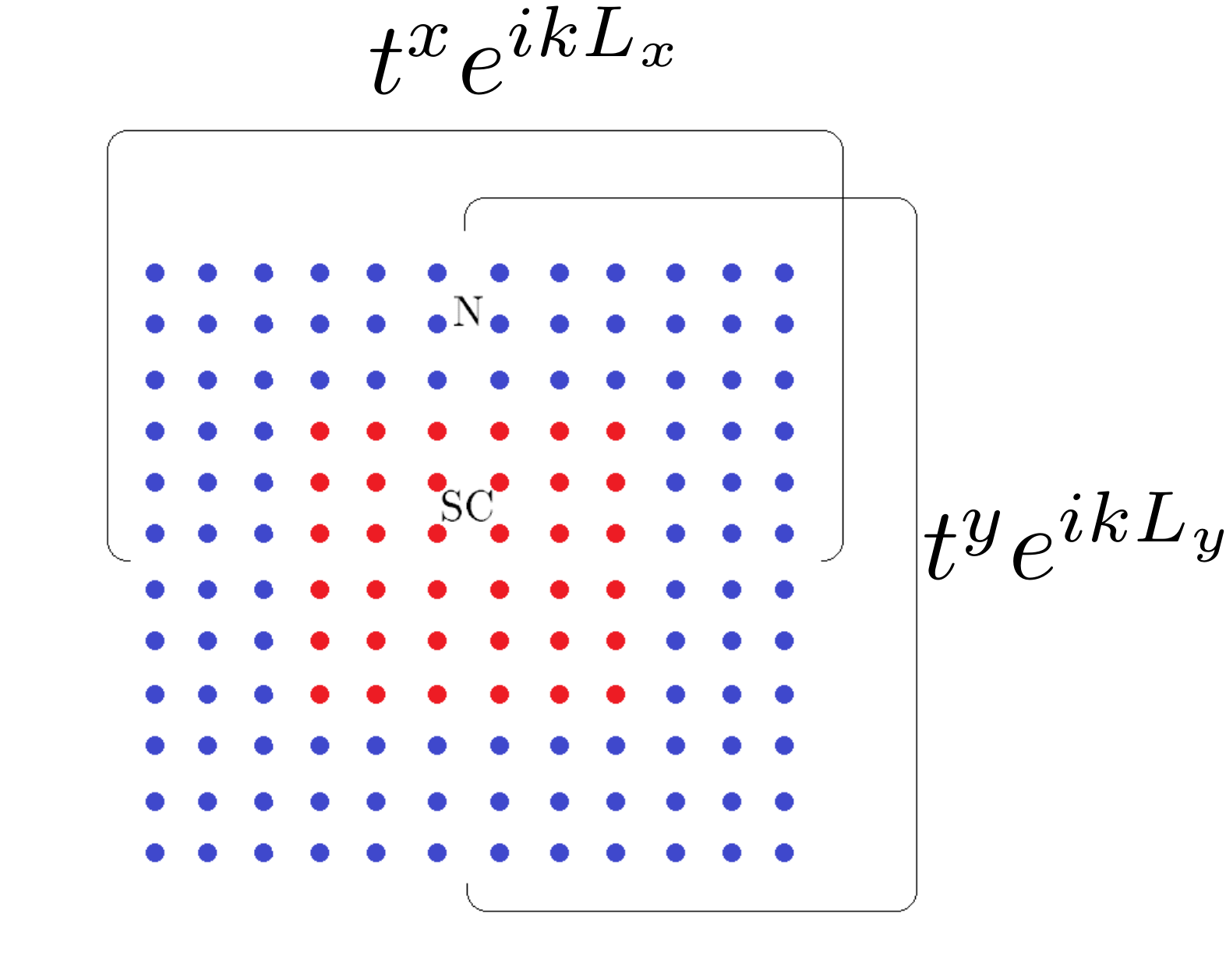}}

\caption{An illustration of a tight-binding unit cell and the momenta corresponding
to the super lattice periodicity. Red sites represent regions with
induced SC due to contact with a SC, blue sites represent normal regions
not in proximity to a SC. (a) A one-dimensional tight-binding model
of the super-lattice unit cell. The additional hopping term between
the first and last sites, which rises from Eq.~\eqref{eq: superlattice k space H},
represents the hopping between adjacent unit cells that receives a
$te^{ikL}$ factor. (b) A two-dimensional tight-binding model of the
super-lattice unit cell. In this case, each site located on one of
the edges has a hopping term to the opposite edge, which represents
the hopping between adjacent unit cells and is multiplied by the corresponding
$t^{j}e^{ikL_{j}}$ factor, $j\in\{x,y\}$ being the hopping direction. \label{fig: Momentum space superlattice unit cell}}
\end{figure}

We utilize the periodic nature of the super-lattice potential
in order to analyze the topological properties of our tight-binding
Hamiltonians. To do so, let us write a general tight-binding model
for a periodic system:
\begin{equation}
H=\underset{rr'jj'}{\sum}c\dag_{r,j}h\left(r-r',j,j'\right)c_{r',j'},\label{eq: real space superlattice k space H}
\end{equation}
where the $r,r'$ indices represent the unit cell index and the $j,j'$
indices represent the sites inside each unit cell including all of
their degrees of freedom (site index, spin, particle-hole). The matrix
element $h$ depends only on the difference $R=r-r'$ due to the periodic
nature of the super-lattice. The Hamiltonian in the space of momentum
which corresponds to the super lattice unit cells can be written as:
\begin{equation}
\begin{split}
&H=\underset{k,j,j'}{\sum}c\dag_{k,j}h_{k}\left(j,j'\right)c_{k,j'},\\
&h_{k}\left(j,j'\right)=\sum_{R=-\inf}^{\inf}e^{ikR}h\left(R,j,j'\right),\\
&c\dag_{r,j}=\underset{k}{\sum}c\dag_{k,j}e^{ikr}.
\end{split}\label{eq: superlattice k space H}
\end{equation}

For nearest neighbors hopping models such as the one presented above,
$h\left(R,j,j'\right)$ is non-zero only for $R=0,\pm1$. The information
describing the internal structure of the unit cell will reside in
$h\left(0,j,j'\right)$, and the hopping between neighboring unit
cells elements will be in \textbf{$h\left(\pm1,j,j'\right)$.} If $n^*$ is the number of sites composing a unit cell, $k$ runs between $-\pi/(n^*a)$ and
$\pi/(n^*a)$, while the usual tight-binding momentum defined for uniform systems runs between $-\pi/a$ and $\pi/a$.  Thus, the tight-binding equivalent
of the $\mathbb{Z}_{2}$ topological invariant given in Eq. \eqref{eq:Kiteav invariant}
would be:
\begin{equation}
\mathcal{Q}={\rm sgn}\left\{ {\rm Pf}\left[\Lambda H(k=0)\right]\right\} \cdot{\rm sgn}\left\{ {\rm Pf}\left[\Lambda H(k=\frac{\pi}{L})\right]\right\}, \label{eq:Tight binding invariant}
\end{equation}with $\P=\Lambda\kappa$ the particle-hole operator and $\Lambda$ is given by:
\[
\Lambda=1_{\rm sites}\tens i\tau_{y}\tens\s_{y},
\]
with $1_{\rm sites}$ the Identity matrix in the space of tight-binding
sites.

The extension of this super-lattice momentum space treatment to two
dimensions is similar to the one-dimensional case shown in Eqs. \eqref{eq: real space superlattice k space H}
and \eqref{eq: superlattice k space H}, resulting in a two-dimensional
momentum space Hamiltonian $h_{{\bf k}}$. If $\psi_j({\bf k})$ are the
eigenstates of $h_{{\bf k}}$, the $\mathbb{Z}$ topological invariant
is given by the Chern number:
\begin{equation}\label{eq:Chern_number}
\nu=\frac{1}{\pi}\underset{j}{\sum}\int_{-\frac{\pi}{L_x}}^{\frac{\pi}{L_x}} dk_{x}\int_{-\frac{\pi}{L_y}}^{\frac{\pi}{L_y}}dk_{y}{\rm Im}\braket{\partial_{k_{x}}\psi_j({\bf k})}{\partial_{k_{y}}\psi_j({\bf k})},
\end{equation}
where the sum runs over the negative energy bands. An illustration of
the super-lattice unit cell and momentum is presented in Fig. \ref{fig: Momentum space superlattice unit cell}.

\subsection{Local Density of States}\label{app:ldos}

The LDOS presented in Fig.~\hyperref[fig:1DL<<LAM]{\ref{fig:1DL<<LAM}(d)} is given by:
\begin{equation}
\mathcal{N}(\vec{r},E)=\text{\textminus}\frac{1}{\pi}{\rm Im}[G(\vec{r},\vec{r},E)],\label{eq:LDOS}
\end{equation}
where $G(\vec{r},\vec{r}',E)=\bra{\vec{r}}G^{R}(E)\ket{\vec{r}'}$,
and the Green's function is a matrix obtained by numerically inverting
the BdG Hamiltonian:

\begin{equation}
G^{R}(E)=\underset{\eta\rightarrow0^{+}}{\rm lim}[E+i\eta\text{\textminus}H_{\rm tb}]^{\text{\textminus}1}.\label{eq:retarded greens}
\end{equation}

\bibliography{references_SC_lattice}

\begin{thebibliography}{50}%
\makeatletter
\providecommand \@ifxundefined [1]{%
 \@ifx{#1\undefined}
}%
\providecommand \@ifnum [1]{%
 \ifnum #1\expandafter \@firstoftwo
 \else \expandafter \@secondoftwo
 \fi
}%
\providecommand \@ifx [1]{%
 \ifx #1\expandafter \@firstoftwo
 \else \expandafter \@secondoftwo
 \fi
}%
\providecommand \natexlab [1]{#1}%
\providecommand \enquote  [1]{``#1''}%
\providecommand \bibnamefont  [1]{#1}%
\providecommand \bibfnamefont [1]{#1}%
\providecommand \citenamefont [1]{#1}%
\providecommand \href@noop [0]{\@secondoftwo}%
\providecommand \href [0]{\begingroup \@sanitize@url \@href}%
\providecommand \@href[1]{\@@startlink{#1}\@@href}%
\providecommand \@@href[1]{\endgroup#1\@@endlink}%
\providecommand \@sanitize@url [0]{\catcode `\\12\catcode `\$12\catcode
  `\&12\catcode `\#12\catcode `\^12\catcode `\_12\catcode `\%12\relax}%
\providecommand \@@startlink[1]{}%
\providecommand \@@endlink[0]{}%
\providecommand \url  [0]{\begingroup\@sanitize@url \@url }%
\providecommand \@url [1]{\endgroup\@href {#1}{\urlprefix }}%
\providecommand \urlprefix  [0]{URL }%
\providecommand \Eprint [0]{\href }%
\providecommand \doibase [0]{http://dx.doi.org/}%
\providecommand \selectlanguage [0]{\@gobble}%
\providecommand \bibinfo  [0]{\@secondoftwo}%
\providecommand \bibfield  [0]{\@secondoftwo}%
\providecommand \translation [1]{[#1]}%
\providecommand \BibitemOpen [0]{}%
\providecommand \bibitemStop [0]{}%
\providecommand \bibitemNoStop [0]{.\EOS\space}%
\providecommand \EOS [0]{\spacefactor3000\relax}%
\providecommand \BibitemShut  [1]{\csname bibitem#1\endcsname}%
\let\auto@bib@innerbib\@empty
\bibitem [{\citenamefont {Alicea}(2012)}]{0034-4885-75-7-076501}%
  \BibitemOpen
  \bibfield  {author} {\bibinfo {author} {\bibfnamefont {J.}~\bibnamefont
  {Alicea}},\ }\href {http://stacks.iop.org/0034-4885/75/i=7/a=076501}
  {\bibfield  {journal} {\bibinfo  {journal} {Rep. Prog. Phys.}\ }\textbf
  {\bibinfo {volume} {75}},\ \bibinfo {pages} {076501} (\bibinfo {year}
  {2012})}\BibitemShut {NoStop}%
\bibitem [{\citenamefont {Beenakker}(2013)}]{Beenakker2013search}%
  \BibitemOpen
  \bibfield  {author} {\bibinfo {author} {\bibfnamefont {C.}~\bibnamefont
  {Beenakker}},\ }\href {\doibase 10.1146/annurev-conmatphys-030212-184337}
  {\bibfield  {journal} {\bibinfo  {journal} {Ann. Rev. Condens. Matt. Phys.}\
  }\textbf {\bibinfo {volume} {4}},\ \bibinfo {pages} {113} (\bibinfo {year}
  {2013})}\BibitemShut {NoStop}%
\bibitem [{\citenamefont {Fu}\ and\ \citenamefont
  {Kane}(2008)}]{PhysRevLett.100.096407}%
  \BibitemOpen
  \bibfield  {author} {\bibinfo {author} {\bibfnamefont {L.}~\bibnamefont
  {Fu}}\ and\ \bibinfo {author} {\bibfnamefont {C.~L.}\ \bibnamefont {Kane}},\
  }\href {\doibase 10.1103/PhysRevLett.100.096407} {\bibfield  {journal}
  {\bibinfo  {journal} {Phys. Rev. Lett.}\ }\textbf {\bibinfo {volume} {100}},\
  \bibinfo {pages} {096407} (\bibinfo {year} {2008})}\BibitemShut {NoStop}%
\bibitem [{\citenamefont {Fu}\ and\ \citenamefont
  {Kane}(2009)}]{PhysRevB.79.161408}%
  \BibitemOpen
  \bibfield  {author} {\bibinfo {author} {\bibfnamefont {L.}~\bibnamefont
  {Fu}}\ and\ \bibinfo {author} {\bibfnamefont {C.~L.}\ \bibnamefont {Kane}},\
  }\href {\doibase 10.1103/PhysRevB.79.161408} {\bibfield  {journal} {\bibinfo
  {journal} {Phys. Rev. B}\ }\textbf {\bibinfo {volume} {79}},\ \bibinfo
  {pages} {161408} (\bibinfo {year} {2009})}\BibitemShut {NoStop}%
\bibitem [{\citenamefont {Sau}\ \emph {et~al.}(2010)\citenamefont {Sau},
  \citenamefont {Lutchyn}, \citenamefont {Tewari},\ and\ \citenamefont
  {Das~Sarma}}]{Sau2010}%
  \BibitemOpen
  \bibfield  {author} {\bibinfo {author} {\bibfnamefont {J.~D.}\ \bibnamefont
  {Sau}}, \bibinfo {author} {\bibfnamefont {R.~M.}\ \bibnamefont {Lutchyn}},
  \bibinfo {author} {\bibfnamefont {S.}~\bibnamefont {Tewari}}, \ and\ \bibinfo
  {author} {\bibfnamefont {S.}~\bibnamefont {Das~Sarma}},\ }\href {\doibase
  10.1103/PhysRevLett.104.040502} {\bibfield  {journal} {\bibinfo  {journal}
  {Phys. Rev. Lett.}\ }\textbf {\bibinfo {volume} {104}},\ \bibinfo {pages}
  {040502} (\bibinfo {year} {2010})}\BibitemShut {NoStop}%
\bibitem [{\citenamefont {Alicea}(2010)}]{PhysRevB.81.125318}%
  \BibitemOpen
  \bibfield  {author} {\bibinfo {author} {\bibfnamefont {J.}~\bibnamefont
  {Alicea}},\ }\href {\doibase 10.1103/PhysRevB.81.125318} {\bibfield
  {journal} {\bibinfo  {journal} {Phys. Rev. B}\ }\textbf {\bibinfo {volume}
  {81}},\ \bibinfo {pages} {125318} (\bibinfo {year} {2010})}\BibitemShut
  {NoStop}%
\bibitem [{\citenamefont {Duckheim}\ and\ \citenamefont
  {Brouwer}(2011)}]{Duckheim2011andreev}%
  \BibitemOpen
  \bibfield  {author} {\bibinfo {author} {\bibfnamefont {M.}~\bibnamefont
  {Duckheim}}\ and\ \bibinfo {author} {\bibfnamefont {P.~W.}\ \bibnamefont
  {Brouwer}},\ }\href {\doibase 10.1103/PhysRevB.83.054513} {\bibfield
  {journal} {\bibinfo  {journal} {Phys. Rev. B}\ }\textbf {\bibinfo {volume}
  {83}},\ \bibinfo {pages} {054513} (\bibinfo {year} {2011})}\BibitemShut
  {NoStop}%
\bibitem [{\citenamefont {Lutchyn}\ \emph {et~al.}(2010)\citenamefont
  {Lutchyn}, \citenamefont {Sau},\ and\ \citenamefont
  {Das~Sarma}}]{PhysRevLett.105.077001}%
  \BibitemOpen
  \bibfield  {author} {\bibinfo {author} {\bibfnamefont {R.~M.}\ \bibnamefont
  {Lutchyn}}, \bibinfo {author} {\bibfnamefont {J.~D.}\ \bibnamefont {Sau}}, \
  and\ \bibinfo {author} {\bibfnamefont {S.}~\bibnamefont {Das~Sarma}},\ }\href
  {\doibase 10.1103/PhysRevLett.105.077001} {\bibfield  {journal} {\bibinfo
  {journal} {Phys. Rev. Lett.}\ }\textbf {\bibinfo {volume} {105}},\ \bibinfo
  {pages} {077001} (\bibinfo {year} {2010})}\BibitemShut {NoStop}%
\bibitem [{\citenamefont {Oreg}\ \emph {et~al.}(2010)\citenamefont {Oreg},
  \citenamefont {Refael},\ and\ \citenamefont {von
  Oppen}}]{PhysRevLett.105.177002}%
  \BibitemOpen
  \bibfield  {author} {\bibinfo {author} {\bibfnamefont {Y.}~\bibnamefont
  {Oreg}}, \bibinfo {author} {\bibfnamefont {G.}~\bibnamefont {Refael}}, \ and\
  \bibinfo {author} {\bibfnamefont {F.}~\bibnamefont {von Oppen}},\ }\href
  {\doibase 10.1103/PhysRevLett.105.177002} {\bibfield  {journal} {\bibinfo
  {journal} {Phys. Rev. Lett.}\ }\textbf {\bibinfo {volume} {105}},\ \bibinfo
  {pages} {177002} (\bibinfo {year} {2010})}\BibitemShut {NoStop}%
\bibitem [{\citenamefont {Mourik}\ \emph {et~al.}(2012)\citenamefont {Mourik},
  \citenamefont {Zuo}, \citenamefont {Frolov}, \citenamefont {Plissard},
  \citenamefont {Bakkers},\ and\ \citenamefont {Kouwenhoven}}]{Mourik1003}%
  \BibitemOpen
  \bibfield  {author} {\bibinfo {author} {\bibfnamefont {V.}~\bibnamefont
  {Mourik}}, \bibinfo {author} {\bibfnamefont {K.}~\bibnamefont {Zuo}},
  \bibinfo {author} {\bibfnamefont {S.~M.}\ \bibnamefont {Frolov}}, \bibinfo
  {author} {\bibfnamefont {S.~R.}\ \bibnamefont {Plissard}}, \bibinfo {author}
  {\bibfnamefont {E.~P. A.~M.}\ \bibnamefont {Bakkers}}, \ and\ \bibinfo
  {author} {\bibfnamefont {L.~P.}\ \bibnamefont {Kouwenhoven}},\ }\href
  {\doibase 10.1126/science.1222360} {\bibfield  {journal} {\bibinfo  {journal}
  {Science}\ }\textbf {\bibinfo {volume} {336}},\ \bibinfo {pages} {1003}
  (\bibinfo {year} {2012})}\BibitemShut {NoStop}%
\bibitem [{\citenamefont {Das}\ \emph {et~al.}(2012)\citenamefont {Das},
  \citenamefont {Ronen}, \citenamefont {Most}, \citenamefont {Oreg},
  \citenamefont {Heiblum},\ and\ \citenamefont {Shtrikman}}]{Das2012}%
  \BibitemOpen
  \bibfield  {author} {\bibinfo {author} {\bibfnamefont {A.}~\bibnamefont
  {Das}}, \bibinfo {author} {\bibfnamefont {Y.}~\bibnamefont {Ronen}}, \bibinfo
  {author} {\bibfnamefont {Y.}~\bibnamefont {Most}}, \bibinfo {author}
  {\bibfnamefont {Y.}~\bibnamefont {Oreg}}, \bibinfo {author} {\bibfnamefont
  {M.}~\bibnamefont {Heiblum}}, \ and\ \bibinfo {author} {\bibfnamefont
  {H.}~\bibnamefont {Shtrikman}},\ }\href {\doibase 10.1038/nphys2479}
  {\bibfield  {journal} {\bibinfo  {journal} {Nature Physics}\ }\textbf
  {\bibinfo {volume} {8}},\ \bibinfo {pages} {887} (\bibinfo {year}
  {2012})}\BibitemShut {NoStop}%
\bibitem [{\citenamefont {Deng}\ \emph {et~al.}(2012)\citenamefont {Deng},
  \citenamefont {Yu}, \citenamefont {Huang}, \citenamefont {Larsson},
  \citenamefont {Caroff},\ and\ \citenamefont {Xu}}]{Deng2012Anomalous}%
  \BibitemOpen
  \bibfield  {author} {\bibinfo {author} {\bibfnamefont {M.~T.}\ \bibnamefont
  {Deng}}, \bibinfo {author} {\bibfnamefont {C.~L.}\ \bibnamefont {Yu}},
  \bibinfo {author} {\bibfnamefont {G.~Y.}\ \bibnamefont {Huang}}, \bibinfo
  {author} {\bibfnamefont {M.}~\bibnamefont {Larsson}}, \bibinfo {author}
  {\bibfnamefont {P.}~\bibnamefont {Caroff}}, \ and\ \bibinfo {author}
  {\bibfnamefont {H.~Q.}\ \bibnamefont {Xu}},\ }\href {\doibase
  10.1021/nl303758w} {\bibfield  {journal} {\bibinfo  {journal} {Nano Letters}\
  }\textbf {\bibinfo {volume} {12}},\ \bibinfo {pages} {6414} (\bibinfo {year}
  {2012})},\ \bibinfo {note} {pMID: 23181691}\BibitemShut {NoStop}%
\bibitem [{\citenamefont {Rokhinson}\ \emph {et~al.}(2012)\citenamefont
  {Rokhinson}, \citenamefont {Liu},\ and\ \citenamefont
  {Furdyna}}]{Rokhinson2012}%
  \BibitemOpen
  \bibfield  {author} {\bibinfo {author} {\bibfnamefont {L.~P.}\ \bibnamefont
  {Rokhinson}}, \bibinfo {author} {\bibfnamefont {X.}~\bibnamefont {Liu}}, \
  and\ \bibinfo {author} {\bibfnamefont {J.~K.}\ \bibnamefont {Furdyna}},\
  }\href {\doibase 10.1038/nphys2429} {\bibfield  {journal} {\bibinfo
  {journal} {Nature Physics}\ }\textbf {\bibinfo {volume} {8}},\ \bibinfo
  {pages} {795} (\bibinfo {year} {2012})}\BibitemShut {NoStop}%
\bibitem [{\citenamefont {Finck}\ \emph {et~al.}(2013)\citenamefont {Finck},
  \citenamefont {Van~Harlingen}, \citenamefont {Mohseni}, \citenamefont
  {Jung},\ and\ \citenamefont {Li}}]{PhysRevLett.110.126406}%
  \BibitemOpen
  \bibfield  {author} {\bibinfo {author} {\bibfnamefont {A.~D.~K.}\
  \bibnamefont {Finck}}, \bibinfo {author} {\bibfnamefont {D.~J.}\ \bibnamefont
  {Van~Harlingen}}, \bibinfo {author} {\bibfnamefont {P.~K.}\ \bibnamefont
  {Mohseni}}, \bibinfo {author} {\bibfnamefont {K.}~\bibnamefont {Jung}}, \
  and\ \bibinfo {author} {\bibfnamefont {X.}~\bibnamefont {Li}},\ }\href
  {\doibase 10.1103/PhysRevLett.110.126406} {\bibfield  {journal} {\bibinfo
  {journal} {Phys. Rev. Lett.}\ }\textbf {\bibinfo {volume} {110}},\ \bibinfo
  {pages} {126406} (\bibinfo {year} {2013})}\BibitemShut {NoStop}%
\bibitem [{\citenamefont {Churchill}\ \emph {et~al.}(2013)\citenamefont
  {Churchill}, \citenamefont {Fatemi}, \citenamefont {Grove-Rasmussen},
  \citenamefont {Deng}, \citenamefont {Caroff}, \citenamefont {Xu},\ and\
  \citenamefont {Marcus}}]{PhysRevB.87.241401}%
  \BibitemOpen
  \bibfield  {author} {\bibinfo {author} {\bibfnamefont {H.~O.~H.}\
  \bibnamefont {Churchill}}, \bibinfo {author} {\bibfnamefont {V.}~\bibnamefont
  {Fatemi}}, \bibinfo {author} {\bibfnamefont {K.}~\bibnamefont
  {Grove-Rasmussen}}, \bibinfo {author} {\bibfnamefont {M.~T.}\ \bibnamefont
  {Deng}}, \bibinfo {author} {\bibfnamefont {P.}~\bibnamefont {Caroff}},
  \bibinfo {author} {\bibfnamefont {H.~Q.}\ \bibnamefont {Xu}}, \ and\ \bibinfo
  {author} {\bibfnamefont {C.~M.}\ \bibnamefont {Marcus}},\ }\href {\doibase
  10.1103/PhysRevB.87.241401} {\bibfield  {journal} {\bibinfo  {journal} {Phys.
  Rev. B}\ }\textbf {\bibinfo {volume} {87}},\ \bibinfo {pages} {241401}
  (\bibinfo {year} {2013})}\BibitemShut {NoStop}%
\bibitem [{\citenamefont {Albrecht}\ \emph {et~al.}(2016)\citenamefont
  {Albrecht}, \citenamefont {Higginbotham}, \citenamefont {Madsen},
  \citenamefont {Kuemmeth}, \citenamefont {Jespersen}, \citenamefont {Nygard},
  \citenamefont {Krogstrup},\ and\ \citenamefont
  {Marcus}}]{albrecht2016exponential}%
  \BibitemOpen
  \bibfield  {author} {\bibinfo {author} {\bibfnamefont {S.}~\bibnamefont
  {Albrecht}}, \bibinfo {author} {\bibfnamefont {A.}~\bibnamefont
  {Higginbotham}}, \bibinfo {author} {\bibfnamefont {M.}~\bibnamefont
  {Madsen}}, \bibinfo {author} {\bibfnamefont {F.}~\bibnamefont {Kuemmeth}},
  \bibinfo {author} {\bibfnamefont {T.}~\bibnamefont {Jespersen}}, \bibinfo
  {author} {\bibfnamefont {J.}~\bibnamefont {Nygard}}, \bibinfo {author}
  {\bibfnamefont {P.}~\bibnamefont {Krogstrup}}, \ and\ \bibinfo {author}
  {\bibfnamefont {C.}~\bibnamefont {Marcus}},\ }\href
  {http://www.nature.com/nature/journal/v531/n7593/full/nature17162.html?WT.feed_name=subjects_electronics-photonics-and-device-physics}
  {\bibfield  {journal} {\bibinfo  {journal} {Nature}\ }\textbf {\bibinfo
  {volume} {531}},\ \bibinfo {pages} {206} (\bibinfo {year}
  {2016})}\BibitemShut {NoStop}%
\bibitem [{Flo()}]{FloatingSuperconductorComment}%
  \BibitemOpen
  \href@noop {} {}\bibinfo {note} {We note that for a single wire, while this
  configuration allows to probe the topological nature of the system, it does
  not exhibit a protected ground-state degeneracy, since the finite charging
  energy of the system generally splits the even- and odd-fermion-number
  states.}\BibitemShut {Stop}%
\bibitem [{\citenamefont {Nadj-Perge}\ \emph {et~al.}(2014)\citenamefont
  {Nadj-Perge}, \citenamefont {Drozdov}, \citenamefont {Li}, \citenamefont
  {Chen}, \citenamefont {Jeon}, \citenamefont {Seo}, \citenamefont {MacDonald},
  \citenamefont {Bernevig},\ and\ \citenamefont {Yazdani}}]{Nadj-Perge602}%
  \BibitemOpen
  \bibfield  {author} {\bibinfo {author} {\bibfnamefont {S.}~\bibnamefont
  {Nadj-Perge}}, \bibinfo {author} {\bibfnamefont {I.~K.}\ \bibnamefont
  {Drozdov}}, \bibinfo {author} {\bibfnamefont {J.}~\bibnamefont {Li}},
  \bibinfo {author} {\bibfnamefont {H.}~\bibnamefont {Chen}}, \bibinfo {author}
  {\bibfnamefont {S.}~\bibnamefont {Jeon}}, \bibinfo {author} {\bibfnamefont
  {J.}~\bibnamefont {Seo}}, \bibinfo {author} {\bibfnamefont {A.~H.}\
  \bibnamefont {MacDonald}}, \bibinfo {author} {\bibfnamefont {B.~A.}\
  \bibnamefont {Bernevig}}, \ and\ \bibinfo {author} {\bibfnamefont
  {A.}~\bibnamefont {Yazdani}},\ }\href {\doibase 10.1126/science.1259327}
  {\bibfield  {journal} {\bibinfo  {journal} {Science}\ }\textbf {\bibinfo
  {volume} {346}},\ \bibinfo {pages} {602} (\bibinfo {year}
  {2014})}\BibitemShut {NoStop}%
\bibitem [{\citenamefont {Pawlak}\ \emph {et~al.}(2016)\citenamefont {Pawlak},
  \citenamefont {Kisiel}, \citenamefont {Klinovaja}, \citenamefont {Meier},
  \citenamefont {Kawai}, \citenamefont {Glatzel}, \citenamefont {Loss},\ and\
  \citenamefont {Meyer}}]{Pawlak2016probing}%
  \BibitemOpen
  \bibfield  {author} {\bibinfo {author} {\bibfnamefont {R.}~\bibnamefont
  {Pawlak}}, \bibinfo {author} {\bibfnamefont {M.}~\bibnamefont {Kisiel}},
  \bibinfo {author} {\bibfnamefont {J.}~\bibnamefont {Klinovaja}}, \bibinfo
  {author} {\bibfnamefont {T.}~\bibnamefont {Meier}}, \bibinfo {author}
  {\bibfnamefont {S.}~\bibnamefont {Kawai}}, \bibinfo {author} {\bibfnamefont
  {T.}~\bibnamefont {Glatzel}}, \bibinfo {author} {\bibfnamefont
  {D.}~\bibnamefont {Loss}}, \ and\ \bibinfo {author} {\bibfnamefont
  {E.}~\bibnamefont {Meyer}},\ }\href
  {https://www.nature.com/articles/npjqi201635} {\bibfield  {journal} {\bibinfo
   {journal} {NPJ Quantum Information}\ }\textbf {\bibinfo {volume} {2}},\
  \bibinfo {pages} {16035} (\bibinfo {year} {2016})}\BibitemShut {NoStop}%
\bibitem [{\citenamefont {Ruby}\ \emph {et~al.}(2015)\citenamefont {Ruby},
  \citenamefont {Pientka}, \citenamefont {Peng}, \citenamefont {von Oppen},
  \citenamefont {Heinrich},\ and\ \citenamefont {Franke}}]{Ruby2015end}%
  \BibitemOpen
  \bibfield  {author} {\bibinfo {author} {\bibfnamefont {M.}~\bibnamefont
  {Ruby}}, \bibinfo {author} {\bibfnamefont {F.}~\bibnamefont {Pientka}},
  \bibinfo {author} {\bibfnamefont {Y.}~\bibnamefont {Peng}}, \bibinfo {author}
  {\bibfnamefont {F.}~\bibnamefont {von Oppen}}, \bibinfo {author}
  {\bibfnamefont {B.~W.}\ \bibnamefont {Heinrich}}, \ and\ \bibinfo {author}
  {\bibfnamefont {K.~J.}\ \bibnamefont {Franke}},\ }\href {\doibase
  10.1103/PhysRevLett.115.197204} {\bibfield  {journal} {\bibinfo  {journal}
  {Phys. Rev. Lett.}\ }\textbf {\bibinfo {volume} {115}},\ \bibinfo {pages}
  {197204} (\bibinfo {year} {2015})}\BibitemShut {NoStop}%
\bibitem [{\citenamefont {Nadj-Perge}\ \emph {et~al.}(2013)\citenamefont
  {Nadj-Perge}, \citenamefont {Drozdov}, \citenamefont {Bernevig},\ and\
  \citenamefont {Yazdani}}]{Nadj-Perge2013proposal}%
  \BibitemOpen
  \bibfield  {author} {\bibinfo {author} {\bibfnamefont {S.}~\bibnamefont
  {Nadj-Perge}}, \bibinfo {author} {\bibfnamefont {I.~K.}\ \bibnamefont
  {Drozdov}}, \bibinfo {author} {\bibfnamefont {B.~A.}\ \bibnamefont
  {Bernevig}}, \ and\ \bibinfo {author} {\bibfnamefont {A.}~\bibnamefont
  {Yazdani}},\ }\href {\doibase 10.1103/PhysRevB.88.020407} {\bibfield
  {journal} {\bibinfo  {journal} {Phys. Rev. B}\ }\textbf {\bibinfo {volume}
  {88}},\ \bibinfo {pages} {020407} (\bibinfo {year} {2013})}\BibitemShut
  {NoStop}%
\bibitem [{\citenamefont {Braunecker}\ and\ \citenamefont
  {Simon}(2013)}]{Braunecker2013interplay}%
  \BibitemOpen
  \bibfield  {author} {\bibinfo {author} {\bibfnamefont {B.}~\bibnamefont
  {Braunecker}}\ and\ \bibinfo {author} {\bibfnamefont {P.}~\bibnamefont
  {Simon}},\ }\href {\doibase 10.1103/PhysRevLett.111.147202} {\bibfield
  {journal} {\bibinfo  {journal} {Phys. Rev. Lett.}\ }\textbf {\bibinfo
  {volume} {111}},\ \bibinfo {pages} {147202} (\bibinfo {year}
  {2013})}\BibitemShut {NoStop}%
\bibitem [{\citenamefont {Vazifeh}\ and\ \citenamefont
  {Franz}(2013)}]{Vazifeh2013self}%
  \BibitemOpen
  \bibfield  {author} {\bibinfo {author} {\bibfnamefont {M.~M.}\ \bibnamefont
  {Vazifeh}}\ and\ \bibinfo {author} {\bibfnamefont {M.}~\bibnamefont
  {Franz}},\ }\href {\doibase 10.1103/PhysRevLett.111.206802} {\bibfield
  {journal} {\bibinfo  {journal} {Phys. Rev. Lett.}\ }\textbf {\bibinfo
  {volume} {111}},\ \bibinfo {pages} {206802} (\bibinfo {year}
  {2013})}\BibitemShut {NoStop}%
\bibitem [{\citenamefont {Klinovaja}\ \emph {et~al.}(2013)\citenamefont
  {Klinovaja}, \citenamefont {Stano}, \citenamefont {Yazdani},\ and\
  \citenamefont {Loss}}]{Klinovaja2013topological}%
  \BibitemOpen
  \bibfield  {author} {\bibinfo {author} {\bibfnamefont {J.}~\bibnamefont
  {Klinovaja}}, \bibinfo {author} {\bibfnamefont {P.}~\bibnamefont {Stano}},
  \bibinfo {author} {\bibfnamefont {A.}~\bibnamefont {Yazdani}}, \ and\
  \bibinfo {author} {\bibfnamefont {D.}~\bibnamefont {Loss}},\ }\href {\doibase
  10.1103/PhysRevLett.111.186805} {\bibfield  {journal} {\bibinfo  {journal}
  {Phys. Rev. Lett.}\ }\textbf {\bibinfo {volume} {111}},\ \bibinfo {pages}
  {186805} (\bibinfo {year} {2013})}\BibitemShut {NoStop}%
\bibitem [{\citenamefont {Pientka}\ \emph {et~al.}(2013)\citenamefont
  {Pientka}, \citenamefont {Glazman},\ and\ \citenamefont {von
  Oppen}}]{Pientka2013Topological}%
  \BibitemOpen
  \bibfield  {author} {\bibinfo {author} {\bibfnamefont {F.}~\bibnamefont
  {Pientka}}, \bibinfo {author} {\bibfnamefont {L.~I.}\ \bibnamefont
  {Glazman}}, \ and\ \bibinfo {author} {\bibfnamefont {F.}~\bibnamefont {von
  Oppen}},\ }\href {\doibase 10.1103/PhysRevB.88.155420} {\bibfield  {journal}
  {\bibinfo  {journal} {Phys. Rev. B}\ }\textbf {\bibinfo {volume} {88}},\
  \bibinfo {pages} {155420} (\bibinfo {year} {2013})}\BibitemShut {NoStop}%
\bibitem [{\citenamefont {Brydon}\ \emph {et~al.}(2015)\citenamefont {Brydon},
  \citenamefont {Das~Sarma}, \citenamefont {Hui},\ and\ \citenamefont
  {Sau}}]{Brydon2015topological}%
  \BibitemOpen
  \bibfield  {author} {\bibinfo {author} {\bibfnamefont {P.~M.~R.}\
  \bibnamefont {Brydon}}, \bibinfo {author} {\bibfnamefont {S.}~\bibnamefont
  {Das~Sarma}}, \bibinfo {author} {\bibfnamefont {H.-Y.}\ \bibnamefont {Hui}},
  \ and\ \bibinfo {author} {\bibfnamefont {J.~D.}\ \bibnamefont {Sau}},\ }\href
  {\doibase 10.1103/PhysRevB.91.064505} {\bibfield  {journal} {\bibinfo
  {journal} {Phys. Rev. B}\ }\textbf {\bibinfo {volume} {91}},\ \bibinfo
  {pages} {064505} (\bibinfo {year} {2015})}\BibitemShut {NoStop}%
\bibitem [{\citenamefont {Peng}\ \emph {et~al.}(2015)\citenamefont {Peng},
  \citenamefont {Pientka}, \citenamefont {Glazman},\ and\ \citenamefont {von
  Oppen}}]{Peng2015strong}%
  \BibitemOpen
  \bibfield  {author} {\bibinfo {author} {\bibfnamefont {Y.}~\bibnamefont
  {Peng}}, \bibinfo {author} {\bibfnamefont {F.}~\bibnamefont {Pientka}},
  \bibinfo {author} {\bibfnamefont {L.~I.}\ \bibnamefont {Glazman}}, \ and\
  \bibinfo {author} {\bibfnamefont {F.}~\bibnamefont {von Oppen}},\ }\href
  {\doibase 10.1103/PhysRevLett.114.106801} {\bibfield  {journal} {\bibinfo
  {journal} {Phys. Rev. Lett.}\ }\textbf {\bibinfo {volume} {114}},\ \bibinfo
  {pages} {106801} (\bibinfo {year} {2015})}\BibitemShut {NoStop}%
\bibitem [{\citenamefont {Dumitrescu}\ \emph {et~al.}(2015)\citenamefont
  {Dumitrescu}, \citenamefont {Roberts}, \citenamefont {Tewari}, \citenamefont
  {Sau},\ and\ \citenamefont {Das~Sarma}}]{Dumitrescu2015majorana}%
  \BibitemOpen
  \bibfield  {author} {\bibinfo {author} {\bibfnamefont {E.}~\bibnamefont
  {Dumitrescu}}, \bibinfo {author} {\bibfnamefont {B.}~\bibnamefont {Roberts}},
  \bibinfo {author} {\bibfnamefont {S.}~\bibnamefont {Tewari}}, \bibinfo
  {author} {\bibfnamefont {J.~D.}\ \bibnamefont {Sau}}, \ and\ \bibinfo
  {author} {\bibfnamefont {S.}~\bibnamefont {Das~Sarma}},\ }\href {\doibase
  10.1103/PhysRevB.91.094505} {\bibfield  {journal} {\bibinfo  {journal} {Phys.
  Rev. B}\ }\textbf {\bibinfo {volume} {91}},\ \bibinfo {pages} {094505}
  (\bibinfo {year} {2015})}\BibitemShut {NoStop}%
\bibitem [{\citenamefont {Sau}\ and\ \citenamefont
  {Sarma}(2012)}]{sau2012realizing}%
  \BibitemOpen
  \bibfield  {author} {\bibinfo {author} {\bibfnamefont {J.~D.}\ \bibnamefont
  {Sau}}\ and\ \bibinfo {author} {\bibfnamefont {S.~D.}\ \bibnamefont
  {Sarma}},\ }\href@noop {} {\bibfield  {journal} {\bibinfo  {journal} {Nature
  communications}\ }\textbf {\bibinfo {volume} {3}},\ \bibinfo {pages} {964}
  (\bibinfo {year} {2012})}\BibitemShut {NoStop}%
\bibitem [{\citenamefont {Sau}\ \emph {et~al.}(2012)\citenamefont {Sau},
  \citenamefont {Lin}, \citenamefont {Hui},\ and\ \citenamefont
  {Das~Sarma}}]{sau2012avoidance}%
  \BibitemOpen
  \bibfield  {author} {\bibinfo {author} {\bibfnamefont {J.~D.}\ \bibnamefont
  {Sau}}, \bibinfo {author} {\bibfnamefont {C.~H.}\ \bibnamefont {Lin}},
  \bibinfo {author} {\bibfnamefont {H.-Y.}\ \bibnamefont {Hui}}, \ and\
  \bibinfo {author} {\bibfnamefont {S.}~\bibnamefont {Das~Sarma}},\ }\href
  {\doibase 10.1103/PhysRevLett.108.067001} {\bibfield  {journal} {\bibinfo
  {journal} {Phys. Rev. Lett.}\ }\textbf {\bibinfo {volume} {108}},\ \bibinfo
  {pages} {067001} (\bibinfo {year} {2012})}\BibitemShut {NoStop}%
\bibitem [{\citenamefont {Fulga}\ \emph {et~al.}(2013)\citenamefont {Fulga},
  \citenamefont {Haim}, \citenamefont {Akhmerov},\ and\ \citenamefont
  {Oreg}}]{fulga2013adaptive}%
  \BibitemOpen
  \bibfield  {author} {\bibinfo {author} {\bibfnamefont {I.~C.}\ \bibnamefont
  {Fulga}}, \bibinfo {author} {\bibfnamefont {A.}~\bibnamefont {Haim}},
  \bibinfo {author} {\bibfnamefont {A.~R.}\ \bibnamefont {Akhmerov}}, \ and\
  \bibinfo {author} {\bibfnamefont {Y.}~\bibnamefont {Oreg}},\ }\href
  {http://stacks.iop.org/1367-2630/15/i=4/a=045020} {\bibfield  {journal}
  {\bibinfo  {journal} {New Journal of Physics}\ }\textbf {\bibinfo {volume}
  {15}},\ \bibinfo {pages} {045020} (\bibinfo {year} {2013})}\BibitemShut
  {NoStop}%
\bibitem [{\citenamefont {Malard}\ \emph {et~al.}(2016)\citenamefont {Malard},
  \citenamefont {Japaridze},\ and\ \citenamefont
  {Johannesson}}]{malard2016synthesizing}%
  \BibitemOpen
  \bibfield  {author} {\bibinfo {author} {\bibfnamefont {M.}~\bibnamefont
  {Malard}}, \bibinfo {author} {\bibfnamefont {G.~I.}\ \bibnamefont
  {Japaridze}}, \ and\ \bibinfo {author} {\bibfnamefont {H.}~\bibnamefont
  {Johannesson}},\ }\href@noop {} {\bibfield  {journal} {\bibinfo  {journal}
  {Physical Review B}\ }\textbf {\bibinfo {volume} {94}},\ \bibinfo {pages}
  {115128} (\bibinfo {year} {2016})}\BibitemShut {NoStop}%
\bibitem [{\citenamefont {Hoffman}\ \emph {et~al.}(2016)\citenamefont
  {Hoffman}, \citenamefont {Klinovaja},\ and\ \citenamefont
  {Loss}}]{hoffman2016topological}%
  \BibitemOpen
  \bibfield  {author} {\bibinfo {author} {\bibfnamefont {S.}~\bibnamefont
  {Hoffman}}, \bibinfo {author} {\bibfnamefont {J.}~\bibnamefont {Klinovaja}},
  \ and\ \bibinfo {author} {\bibfnamefont {D.}~\bibnamefont {Loss}},\
  }\href@noop {} {\bibfield  {journal} {\bibinfo  {journal} {Physical Review
  B}\ }\textbf {\bibinfo {volume} {93}},\ \bibinfo {pages} {165418} (\bibinfo
  {year} {2016})}\BibitemShut {NoStop}%
\bibitem [{\citenamefont {Zhang}\ and\ \citenamefont
  {Nori}(2016)}]{zhang2016majorana}%
  \BibitemOpen
  \bibfield  {author} {\bibinfo {author} {\bibfnamefont {P.}~\bibnamefont
  {Zhang}}\ and\ \bibinfo {author} {\bibfnamefont {F.}~\bibnamefont {Nori}},\
  }\href@noop {} {\bibfield  {journal} {\bibinfo  {journal} {New Journal of
  Physics}\ }\textbf {\bibinfo {volume} {18}},\ \bibinfo {pages} {043033}
  (\bibinfo {year} {2016})}\BibitemShut {NoStop}%
\bibitem [{\citenamefont {Lu}\ \emph {et~al.}(2016)\citenamefont {Lu},
  \citenamefont {He}, \citenamefont {Xu}, \citenamefont {Lin},\ and\
  \citenamefont {Law}}]{PhysRevB.94.024507}%
  \BibitemOpen
  \bibfield  {author} {\bibinfo {author} {\bibfnamefont {Y.}~\bibnamefont
  {Lu}}, \bibinfo {author} {\bibfnamefont {W.-Y.}\ \bibnamefont {He}}, \bibinfo
  {author} {\bibfnamefont {D.-H.}\ \bibnamefont {Xu}}, \bibinfo {author}
  {\bibfnamefont {N.}~\bibnamefont {Lin}}, \ and\ \bibinfo {author}
  {\bibfnamefont {K.~T.}\ \bibnamefont {Law}},\ }\href {\doibase
  10.1103/PhysRevB.94.024507} {\bibfield  {journal} {\bibinfo  {journal} {Phys.
  Rev. B}\ }\textbf {\bibinfo {volume} {94}},\ \bibinfo {pages} {024507}
  (\bibinfo {year} {2016})}\BibitemShut {NoStop}%
\bibitem [{\citenamefont {Lee}\ \emph {et~al.}(2012)\citenamefont {Lee},
  \citenamefont {Alicea},\ and\ \citenamefont {Refael}}]{Lee2012electrical}%
  \BibitemOpen
  \bibfield  {author} {\bibinfo {author} {\bibfnamefont {S.-P.}\ \bibnamefont
  {Lee}}, \bibinfo {author} {\bibfnamefont {J.}~\bibnamefont {Alicea}}, \ and\
  \bibinfo {author} {\bibfnamefont {G.}~\bibnamefont {Refael}},\ }\href
  {\doibase 10.1103/PhysRevLett.109.126403} {\bibfield  {journal} {\bibinfo
  {journal} {Phys. Rev. Lett.}\ }\textbf {\bibinfo {volume} {109}},\ \bibinfo
  {pages} {126403} (\bibinfo {year} {2012})}\BibitemShut {NoStop}%
\bibitem [{\citenamefont {Motrunich}\ \emph {et~al.}(2001)\citenamefont
  {Motrunich}, \citenamefont {Damle},\ and\ \citenamefont
  {Huse}}]{Motrunich2001Griffiths}%
  \BibitemOpen
  \bibfield  {author} {\bibinfo {author} {\bibfnamefont {O.}~\bibnamefont
  {Motrunich}}, \bibinfo {author} {\bibfnamefont {K.}~\bibnamefont {Damle}}, \
  and\ \bibinfo {author} {\bibfnamefont {D.~A.}\ \bibnamefont {Huse}},\ }\href
  {\doibase 10.1103/PhysRevB.63.224204} {\bibfield  {journal} {\bibinfo
  {journal} {Phys. Rev. B}\ }\textbf {\bibinfo {volume} {63}},\ \bibinfo
  {pages} {224204} (\bibinfo {year} {2001})}\BibitemShut {NoStop}%
\bibitem [{\citenamefont {Brouwer}\ \emph {et~al.}(2011)\citenamefont
  {Brouwer}, \citenamefont {Duckheim}, \citenamefont {Romito},\ and\
  \citenamefont {von Oppen}}]{Brouwer2011Probability}%
  \BibitemOpen
  \bibfield  {author} {\bibinfo {author} {\bibfnamefont {P.~W.}\ \bibnamefont
  {Brouwer}}, \bibinfo {author} {\bibfnamefont {M.}~\bibnamefont {Duckheim}},
  \bibinfo {author} {\bibfnamefont {A.}~\bibnamefont {Romito}}, \ and\ \bibinfo
  {author} {\bibfnamefont {F.}~\bibnamefont {von Oppen}},\ }\href {\doibase
  10.1103/PhysRevLett.107.196804} {\bibfield  {journal} {\bibinfo  {journal}
  {Phys. Rev. Lett.}\ }\textbf {\bibinfo {volume} {107}},\ \bibinfo {pages}
  {196804} (\bibinfo {year} {2011})}\BibitemShut {NoStop}%
\bibitem [{\citenamefont {Altland}\ and\ \citenamefont
  {Zirnbauer}(1997)}]{PhysRevB.55.1142}%
  \BibitemOpen
  \bibfield  {author} {\bibinfo {author} {\bibfnamefont {A.}~\bibnamefont
  {Altland}}\ and\ \bibinfo {author} {\bibfnamefont {M.~R.}\ \bibnamefont
  {Zirnbauer}},\ }\href {\doibase 10.1103/PhysRevB.55.1142} {\bibfield
  {journal} {\bibinfo  {journal} {Phys. Rev. B}\ }\textbf {\bibinfo {volume}
  {55}},\ \bibinfo {pages} {1142} (\bibinfo {year} {1997})}\BibitemShut
  {NoStop}%
\bibitem [{\citenamefont {Schnyder}\ \emph {et~al.}(2008)\citenamefont
  {Schnyder}, \citenamefont {Ryu}, \citenamefont {Furusaki},\ and\
  \citenamefont {Ludwig}}]{PhysRevB.78.195125}%
  \BibitemOpen
  \bibfield  {author} {\bibinfo {author} {\bibfnamefont {A.~P.}\ \bibnamefont
  {Schnyder}}, \bibinfo {author} {\bibfnamefont {S.}~\bibnamefont {Ryu}},
  \bibinfo {author} {\bibfnamefont {A.}~\bibnamefont {Furusaki}}, \ and\
  \bibinfo {author} {\bibfnamefont {A.~W.~W.}\ \bibnamefont {Ludwig}},\ }\href
  {\doibase 10.1103/PhysRevB.78.195125} {\bibfield  {journal} {\bibinfo
  {journal} {Phys. Rev. B}\ }\textbf {\bibinfo {volume} {78}},\ \bibinfo
  {pages} {195125} (\bibinfo {year} {2008})}\BibitemShut {NoStop}%
\bibitem [{\citenamefont {Kitaev}(2009)}]{kitaev2009periodic}%
  \BibitemOpen
  \bibfield  {author} {\bibinfo {author} {\bibfnamefont {A.}~\bibnamefont
  {Kitaev}},\ }\href {http://aip.scitation.org/doi/abs/10.1063/1.3149495}
  {\bibfield  {journal} {\bibinfo  {journal} {AIP Conf. Proc.}\ }\textbf
  {\bibinfo {volume} {1134}},\ \bibinfo {pages} {22} (\bibinfo {year}
  {2009})}\BibitemShut {NoStop}%
\bibitem [{Fra()}]{FragileSymmetry}%
  \BibitemOpen
  \href@noop {} {}\bibinfo {note} {It can be broken, for example, by slightly
  rotating the magnetic field, or by going beyond the strictly 1d description
  of the nanowire. This would still leave the system in class D, with a
  $\mathbb{Z}_2$ invariant which we wish to obtain.}\BibitemShut {Stop}%
\bibitem [{\citenamefont {Kitaev}(2001)}]{1063-7869-44-10S-S29}%
  \BibitemOpen
  \bibfield  {author} {\bibinfo {author} {\bibfnamefont {A.~Y.}\ \bibnamefont
  {Kitaev}},\ }\href {http://stacks.iop.org/1063-7869/44/i=10S/a=S29}
  {\bibfield  {journal} {\bibinfo  {journal} {Phys.-Usp.}\ }\textbf {\bibinfo
  {volume} {44}},\ \bibinfo {pages} {131} (\bibinfo {year} {2001})}\BibitemShut
  {NoStop}%
\bibitem [{\citenamefont {Ghosh}\ \emph {et~al.}(2010)\citenamefont {Ghosh},
  \citenamefont {Sau}, \citenamefont {Tewari},\ and\ \citenamefont
  {Das~Sarma}}]{ghosh2010non}%
  \BibitemOpen
  \bibfield  {author} {\bibinfo {author} {\bibfnamefont {P.}~\bibnamefont
  {Ghosh}}, \bibinfo {author} {\bibfnamefont {J.~D.}\ \bibnamefont {Sau}},
  \bibinfo {author} {\bibfnamefont {S.}~\bibnamefont {Tewari}}, \ and\ \bibinfo
  {author} {\bibfnamefont {S.}~\bibnamefont {Das~Sarma}},\ }\href {\doibase
  10.1103/PhysRevB.82.184525} {\bibfield  {journal} {\bibinfo  {journal} {Phys.
  Rev. B}\ }\textbf {\bibinfo {volume} {82}},\ \bibinfo {pages} {184525}
  (\bibinfo {year} {2010})}\BibitemShut {NoStop}%
\bibitem [{\citenamefont {Tewari}\ and\ \citenamefont
  {Sau}(2012)}]{tewari2012topological}%
  \BibitemOpen
  \bibfield  {author} {\bibinfo {author} {\bibfnamefont {S.}~\bibnamefont
  {Tewari}}\ and\ \bibinfo {author} {\bibfnamefont {J.~D.}\ \bibnamefont
  {Sau}},\ }\href
  {https://journals.aps.org/prl/abstract/10.1103/PhysRevLett.109.150408}
  {\bibfield  {journal} {\bibinfo  {journal} {Phys. Rev. Lett.}\ }\textbf
  {\bibinfo {volume} {109}},\ \bibinfo {pages} {150408} (\bibinfo {year}
  {2012})}\BibitemShut {NoStop}%
\bibitem [{Gap()}]{GapRemark}%
  \BibitemOpen
  \href@noop {} {}\bibinfo {note} {For a system uniformally covered by a
  superconductor, the gap near the Fermi momenta is approximately given by
  $\Delta_{\rm ind}/\sqrt{1+V_{\rm Z}^2/4E_{\rm so}^2}$, as long as
  $\Delta_{\rm ind}$ is small enough compared to either $V_{\rm Z}$ or $E_{\rm
  so}$. It is the actual gap of the system far enough from the phase transition
  line as it becomes smaller than the gap at $k=0$.}\BibitemShut {Stop}%
\bibitem [{Dre()}]{Dressel}%
  \BibitemOpen
  \href@noop {} {}\bibinfo {note} {This ratio is chosen as it exhibits a large
  energy gap for a system uniformly covered by a SC, in accordance with
  simulations preformed in Ref.~[\onlinecite{PhysRevB.81.125318}]}\BibitemShut
  {NoStop}%
\bibitem [{\citenamefont {Suominen}\ \emph {et~al.}(2017)\citenamefont
  {Suominen}, \citenamefont {Kjaergaard}, \citenamefont {Hamilton},
  \citenamefont {Shabani}, \citenamefont {Palmstrom}, \citenamefont {Marcus},\
  and\ \citenamefont {Nichele}}]{suominen2017scalable}%
  \BibitemOpen
  \bibfield  {author} {\bibinfo {author} {\bibfnamefont {H.~J.}\ \bibnamefont
  {Suominen}}, \bibinfo {author} {\bibfnamefont {M.}~\bibnamefont
  {Kjaergaard}}, \bibinfo {author} {\bibfnamefont {A.~R.}\ \bibnamefont
  {Hamilton}}, \bibinfo {author} {\bibfnamefont {J.}~\bibnamefont {Shabani}},
  \bibinfo {author} {\bibfnamefont {C.~J.}\ \bibnamefont {Palmstrom}}, \bibinfo
  {author} {\bibfnamefont {C.~M.}\ \bibnamefont {Marcus}}, \ and\ \bibinfo
  {author} {\bibfnamefont {F.}~\bibnamefont {Nichele}},\ }\href@noop {}
  {\bibfield  {journal} {\bibinfo  {journal} {arXiv preprint arXiv:1703.03699}\
  } (\bibinfo {year} {2017})}\BibitemShut {NoStop}%
\bibitem [{\citenamefont {{Barkeshli}}\ and\ \citenamefont
  {{Sau}}(2015)}]{barkeshli2015physical}%
  \BibitemOpen
  \bibfield  {author} {\bibinfo {author} {\bibfnamefont {M.}~\bibnamefont
  {{Barkeshli}}}\ and\ \bibinfo {author} {\bibfnamefont {J.~D.}\ \bibnamefont
  {{Sau}}},\ }\href@noop {} {\bibfield  {journal} {\bibinfo  {journal} {ArXiv
  e-prints}\ } (\bibinfo {year} {2015})},\ \Eprint
  {http://arxiv.org/abs/1509.07135} {arXiv:1509.07135 [cond-mat.mes-hall]}
  \BibitemShut {NoStop}%
\bibitem [{\citenamefont {{Landau}}\ \emph {et~al.}(2016)\citenamefont
  {{Landau}}, \citenamefont {{Plugge}}, \citenamefont {{Sela}}, \citenamefont
  {{Altland}}, \citenamefont {{Albrecht}},\ and\ \citenamefont
  {{Egger}}}]{Landau2016}%
  \BibitemOpen
  \bibfield  {author} {\bibinfo {author} {\bibfnamefont {L.~A.}\ \bibnamefont
  {{Landau}}}, \bibinfo {author} {\bibfnamefont {S.}~\bibnamefont {{Plugge}}},
  \bibinfo {author} {\bibfnamefont {E.}~\bibnamefont {{Sela}}}, \bibinfo
  {author} {\bibfnamefont {A.}~\bibnamefont {{Altland}}}, \bibinfo {author}
  {\bibfnamefont {S.~M.}\ \bibnamefont {{Albrecht}}}, \ and\ \bibinfo {author}
  {\bibfnamefont {R.}~\bibnamefont {{Egger}}},\ }\href {\doibase
  10.1103/PhysRevLett.116.050501} {\bibfield  {journal} {\bibinfo  {journal}
  {Physical Review Letters}\ }\textbf {\bibinfo {volume} {116}},\ \bibinfo
  {eid} {050501} (\bibinfo {year} {2016})},\ \Eprint
  {http://arxiv.org/abs/1509.05345} {arXiv:1509.05345 [cond-mat.mes-hall]}
  \BibitemShut {NoStop}%
\end{thebibliography}%

\end{document}